%% file: main.tex
\documentclass[letterpaper, 10 pt, journal, twoside]{ieeetran} 
\usepackage{todonotes}
\IEEEoverridecommandlockouts                              

\usepackage{graphics} 
\usepackage{epsfig} 
\usepackage{times} 
\usepackage{amsmath} 
\usepackage{amssymb}  
\usepackage{hyperref} 
\usepackage{color}
\usepackage[T1]{fontenc}
\usepackage{algorithm}
\usepackage{algpseudocode}
\usepackage{multirow}
\usepackage{ctable} 
\usepackage{subcaption}
\usepackage{amsfonts}
\usepackage{footnote}
\DeclareMathOperator*{\argmax}{arg\!\max}
\DeclareMathOperator*{\argmin}{arg\!\min}

\title{Interaction-Aware Trajectory Prediction and Planning for Autonomous Vehicles \\ in Forced Merge Scenarios}
 \author{Kaiwen Liu, Nan Li, H. Eric Tseng, Ilya Kolmanovsky, and Anouck Girard
 \thanks{Kaiwen Liu, Nan Li, Ilya Kolmanovsky, and Anouck Girard are with the Department of Aerospace Engineering, University of Michigan, Ann Arbor, 48109 MI,
 USA. {\tt\small \{kwliu,nanli,ilya,anouck\}@umich.edu}}
 \thanks{H. Eric Tseng is with the Ford Motor Company, Dearborn, 48126 MI, USA. {\tt\small htseng@ford.com}}
 }

\begin{document}

\maketitle
\thispagestyle{empty}
\pagestyle{empty}

\input{sections/00_abstract.tex}

\input{sections/01_introduction.tex}

\input{sections/02_problem_formulation.tex}

\input{sections/03_leader_follower_game.tex}

\input{sections/04_decision_making_algorithm.tex}

\input{sections/05_results.tex}
\input{sections/06_conclusion.tex}

\bibliographystyle{ieeetr}
\bibliography{references}

\input{sections/07_biography.tex}

\addtolength{\textheight}{-12cm}  

\end{document}

%% file: sections/00_abstract.tex
\begin{abstract}
Merging is, in general, a challenging task for both human drivers and autonomous vehicles, especially in dense traffic, because the merging vehicle typically needs to interact with other vehicles to identify or create a gap and safely merge into. In this paper, we consider the problem of autonomous vehicle control for forced merge scenarios. We propose a novel game-theoretic controller, called the Leader-Follower Game Controller (LFGC), in which the interactions between the autonomous ego vehicle and other vehicles with a priori uncertain driving intentions is modeled as a partially observable leader-follower game. The LFGC estimates the other vehicles' intentions online based on observed trajectories, and then predicts their future trajectories and plans the ego vehicle's own trajectory using Model Predictive Control (MPC) to simultaneously achieve probabilistically guaranteed safety and merging objectives. To verify the performance of LFGC, we test it in simulations and with the NGSIM data, where the LFGC demonstrates a high success rate of $97.5\%$ in merging.
\end{abstract}

%% file: sections/01_introduction.tex
\section{Introduction}

Advances in autonomous vehicle technologies are projected to reduce vehicle crashes and fatalities, improve mobility especially for elderly and disabled people, improve fuel economy and emission control, and to promote more efficient land uses \cite{anderson2014autonomous,meyer2017autonomous, ersal2020connected}. Despite these benefits, there are still many challenges that need to be addressed to deliver a highly (level~4 or level~5) autonomous vehicle \cite{sae2014taxonomy}. One challenging scenario for both human drivers and autonomous vehicles is highway forced merge, where the merging vehicle needs to choose a proper gap in the highway traffic and potentially force the upstream traffic to slow down so that it can safely merge into that gap. Forced merge typically occurs in mandatory merge scenarios where the current lane is ending, such as at highway on-ramps. When the traffic is dense, interactions and/or cooperation between the merging vehicle and vehicles driving in the target lane are often needed. In particular, a vehicle in the target lane may choose to ignore the merging vehicle (i.e., {\it proceed}) and consequently the merging vehicle can only merge behind. Alternatively, the vehicle in the target lane may choose to {\it yield} to the merging vehicle (i.e., let the merging vehicle merge in front of it). In order to successfully merge into a busy traffic, an autonomous vehicle controller needs to appropriately respond to the intentions to proceed or yield of other vehicles. An overly conservative controller may yield to all other vehicles (including those that intend to yield to the autonomous ego vehicle) and eventually fail to merge, while an overly aggressive controller may have conflicts with the vehicles that intend to proceed and lead to vehicle crashes. Meanwhile, the decision whether to proceed or to yield to another vehicle depends not only on the traffic situation (e.g., the relative position and velocity between the two vehicles) but also on its driver's general driving style, personality, mood, etc. For instance, in a similar situation, an aggressive driver may be inclined to proceed while a cautious/conservative driver may tend to yield. This poses a significant challenge to autonomous vehicle planning and control.

\begin{figure}[t]
\begin{center}
\begin{picture}(240.0, 95.0)
\put(  -10,  -10){\epsfig{file=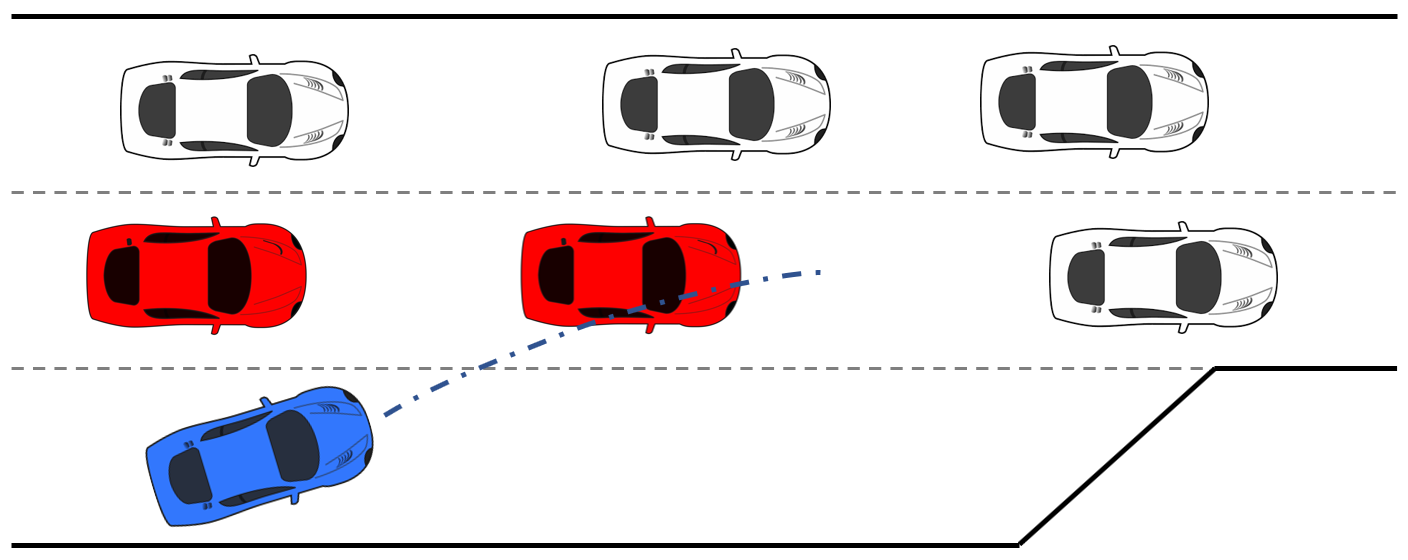,width=1\linewidth}}  
\end{picture}
\end{center}
      \caption{The autonomous vehicle (blue) needs to merge onto the highway before the on-ramp section ends. In dense traffic, there may not be a sufficient gap for the autonomous vehicle to merge into. In this case, the autonomous vehicle needs to force the other vehicles to cooperate and let it cut in. However, interacting vehicles that are aware of the autonomous vehicle's merging attempt (red) may choose to {\it proceed} or {\it yield} depending on their intentions.}
      \label{fig:intro_merge}
\end{figure}

There exists an extensive literature on modeling human driver interactions and autonomous vehicle decision-making during lane change or merging. To handle interaction uncertainties (e.g., due to varied cooperation intentions of other vehicles), the Partially Observable Markov Decision Process (POMDP) framework has been exploited, where the uncertainties are modeled as latent variables and estimated online based on observed trajectories \cite{bandyopadhyay2013intention,dong2017interactive,hubmann2018automated,li2019stochastic,li2019decision,lefkopoulos2021interaction}. However, solving a POMDP problem with a large state and/or action space is computationally very demanding \cite{monahan1982state}. Consequently, conventional POMDP-based approaches typically only consider the interaction of the ego vehicle with one interacting vehicle at a time to minimize the state space dimension. However, in reality a merging scenario can involve simultaneous interactions with multiple vehicles.

Reinforcement Learning (RL) is another popular approach to developing control policies for lane change or merge scenarios \cite{kiran2021deep, baheri2020deep}. An RL-based policy can account for the vehicle interactions in certain scenarios through training in an environment capable of representing such interactions \cite{li2020energy,saxena2020driving}. In order to obtain through RL driving policies that behave like human drivers, several researchers chose to use inverse RL to estimate human's reward function for driving \cite{wu2020efficient, huang2020modeling, rosbach2020planning, menner2020inverse}. To be able to model different human driver styles and/or interaction intentions, \cite{bouton2019cooperation} incorporates cooperativeness into the intelligent driver model and \cite{hu2019interaction} formulates different reward functions for different drivers and performs RL based on the models. Although RL-based approaches are appealing in terms of their potentials to handle complex traffic scenarios with multi-vehicle interactions, potential drawbacks of these approaches that hinder their practical application include their lacks of interpretability and explicit safety guarantee, because safety is typically only promoted through certain terms in the reward function rather than enforced through hard constraints.

To achieve more interpretable control, other researchers proposed to explicitly incorporate a prediction model for vehicle interactions in the control algorithm. For instance, \cite{bae2020cooperation} uses a ``Social Generative Adversarial Network (Social GAN)'' to generate predictions of other vehicles' future trajectories in response to ego vehicle's actions. However, the Social GAN does not account for variations of drivers' styles and intentions and needs to be trained with sufficient traffic data \cite{gupta2018social}. For the latter, it has been reported that multi-vehicle interaction scenarios in released traffic datasets are insufficient \cite{wang2017much}. Game-theoretic methods have also been investigated for modeling vehicle interactions in lane change or merge scenarios \cite{li2019decision, sadigh2016planning, li2017game, fisac2019hierarchical, dai2020towards, schwarting2019social}. It is possible to account for varied driving styles and/or intentions with these game-theoretic methods, for instance, through modeling and online estimation of drivers' cognitive levels \cite{li2017game} or aggressiveness \cite{yu2018human,zhang2019game}.

In this paper, we propose a novel high-level control algorithm, called the Leader-Follower Game Controller (LFGC), for autonomous vehicle planning and control in forced merge scenarios. In the LFGC, drivers' interaction intentions (to {\it proceed} or {\it yield}) and their resulting vehicle behaviors are represented by an explicit game-theoretic model with multiple concurrent leader-follower pairs, called a leader-follower game \cite{nan2021gametheoretic}. To account for interaction uncertainties, the pairwise leader-follower relationships among the vehicles are assumed to be {\it a priori} uncertain and modeled as latent variables. The LFGC estimates the leader-follower relationships online based on observed trajectories and makes optimal decisions for the autonomous ego vehicle using a Model Predictive Control (MPC)-based strategy. The proposed approach thus adapts to the inferred leader-follower relationship estimates to simultaneously achieve probabilistically guaranteed safety and the merging objectives. Note that a similar idea has been investigated in our previous conference paper \cite{liu2021cooperation}. The LFGC presented in this paper differs from the one in \cite{liu2021cooperation} in several aspects: 1) Instead of relying on discretization of the state space and the POMDP framework in \cite{liu2021cooperation}, the LFGC in this paper is designed assuming a continuous state space, which results in smoother trajectories for lower-level controllers to track and which alleviates the computational difficulty associated with discrete spaces. 2) Unlike our previous work \cite{liu2021cooperation} in which we use a small number of actions (or motion primitives) to represent vehicle behavior, the LFGC of this paper predicts and plans vehicle motion using two much larger sets of trajectories ($162$ trajectories for the merging ego vehicle and $81$ trajectories for each of the highway interacting vehicles), which leads to finer-resolution controls and the potential for higher performance. 3) The LFGC of this paper is validated based on a comprehensive set of simulation-based test cases including cases where other vehicles are controlled by various types of driver models and cases where their motion follows real traffic data, which is not done in \cite{liu2021cooperation}. 

The contributions and novelties of the proposed LFGC over aforementioned previous approaches are as follows:
\begin{enumerate}
    \item The LFGC uses a game-theoretic model for vehicle trajectory prediction while accounting for interactions and cooperation and while leading to interpretable control solutions (because the control solutions are based on model predictive control with an interpretable game-theoretic prediction model) .
    \item The LFGC handles interaction uncertainties due to varied cooperation intentions of other vehicles by modeling these uncertainties as latent variables and estimating them online based on observed trajectories and Bayesian inference.
    \item The LFGC represents vehicle safety requirements (e.g., collision avoidance) as constraints and pursues optimization subject to satisfying an explicit probabilistic safety characterization (i.e., a user-specified probability bound of safety) in the presence of interaction uncertainties.
    \item The LFGC is designed in a continuous state space setting, which avoids the computational difficulty resulting from space discretization of previous POMDP-based approaches. This also enables the LFGC to handle more complex scenarios that involve interactions with multiple vehicles.
    \item The LFGC is validated based on a comprehensive set of simulation-based case studies that include cases where other vehicles are controlled by various types of driver models and cases where their motion follows actual vehicle trajectories in the NGSIM US Highway 101 dataset \cite{us101data}. For the latter, the LFGC demonstrates a high success rate (in terms of safely completing merges) of $97.5\%$.
\end{enumerate}


This paper is organized as follows: In Section~\ref{sec:prob_form}, we introduce the models that represent vehicle/traffic dynamics, driver objectives, vehicle actions, and the MPC-based control strategy for the autonomous ego vehicle. In Section~\ref{sec:lf_game}, we introduce the leader-follower game model that is used to represent drivers' interaction intentions and their resulting vehicle behaviors in multi-vehicle traffic scenarios. In Section~\ref{sec:game_decision_making}, we integrate the MPC-based control strategy with the leader-follower game model and with online estimation of pairwise leader-follower relationships among interacting vehicles based on Bayesian inference, to enable the ego vehicle's actions to adapt in real-time to interacting drivers'/vehicles' intentions. In Section~\ref{sec:results}, we validate the proposed LFGC through multiple simulation-based case studies, including validations against vehicles following our leader-follower game model, the Intelligent Driver Model (IDM), and trajectories from NGSIM US Highway 101 data. Finally, conclusions are given in Section~\ref{sec:summary}.

%% file: sections/02_problem_formulation.tex
\section{Models and Control Strategy Descriptions} \label{sec:prob_form}

In this section, we introduce models to represent the vehicle and traffic dynamics and the MPC-based strategy for ego vehicle's trajectory planning.

\subsection{Vehicle dynamics}

We use the kinematic bicycle model \cite{rajamani2011vehicle} to represent the motion of each vehicle. The kinematic bicycle model is defined by the following set of continuous-time equations, 
\begin{equation}\label{eq:kinematic_bicycle}
    \begin{aligned}
    &\dot{x} = v \cos(\psi + \beta), \\
    &\dot{y} = v \sin(\psi + \beta), \\
    &\dot{v} = a, \\
    &\dot{\psi} = \frac{v}{l_r} \sin(\beta), \\
    &\beta = \tan^{-1} \left(\frac{l_r }{l_r + l_f } \tan \delta_f\right),
    \end{aligned}
\end{equation}
where we have assumed only front-wheel steering $\delta_f$ and no rear-wheel steering (i.e., $\delta_r = 0$); $x$ and $y$ are the longitudinal and lateral positions of the vehicle; $v$ is the speed of the vehicle; $\psi$ and $\beta$ are the yaw angle and the slip angle of the vehicle; $l_f$ and $l_r$ represent the distances from the CG of the vehicle to the front wheel and rear wheel axles; $a$ is the acceleration along the direction of speed $v$. The control inputs are the acceleration and front-wheel steering, $u = [a, \delta_f]^T$.

While vehicle models other than \eqref{eq:kinematic_bicycle} could be used, the above kinematic bicycle model \eqref{eq:kinematic_bicycle} is suitable for our purpose of trajectory prediction and planning in forced merge scenarios -- it can produce sufficiently accurate predictions of vehicle trajectories under given acceleration and front-wheel steering profiles \cite{kong2015kinematic} while it is simple and thus computationally efficient.

\subsection{Traffic dynamics}

We consider a traffic scenario involving $n+1$ vehicles, including the ego vehicle, denoted by $0$, and $n$ other interacting vehicles $k$, $k \in \{1,\dots, n\}$, which correspond to vehicles that are aware of the ego vehicle's merging attempt. Therefore, the traffic state and its dynamics are characterized by the aggregation of all $n+1$ vehicles' states and dynamics. Specifically, we describe the traffic dynamics using the following discrete-time model, 
\begin{equation}\label{eq:sys_dyn}
    \bar{s}_{t+1} = f(\bar{s}_t, \bar{u}_t),
\end{equation}
where $\bar{s}_t = (s_t^0, s_t^1, s_t^2, \dots, s_t^n)$ denotes the traffic state at the discrete time instant $t$, with $s_t^0$ denoting the ego vehicle's state and $s_t^k$, $k \in \{1,\dots, n\}$, denoting the $k$th interacting vehicle's state; and similarly, $\bar{u}_t = (u_t^0, u_t^1, u_t^2, \dots, u_t^n)$ denotes the aggregation of all $n+1$ vehicles' control inputs at the time constant $t$. In particular, each vehicle's state $s_t^k$, $k \in \{0, 1, \dots, n\}$, consists of its $x$ and $y$ positions, speed, and yaw angle, i.e., $s_t^k = [x_t^k, y_t^k, v_t^k, \psi_t^k]^T$, and each vehicle's control inputs are $u_t^k = [a_t^k, \delta_{f,t}^k]^T$. Accordingly, the function $f$ in \eqref{eq:sys_dyn} that represents the transition of traffic state from $\bar{s}_t$ to $\bar{s}_{t+1}$ as a result of all vehicles' control inputs $\bar{u}_t$ is an aggregation of $(n+1)$-copies of the kinematic bicycle model \eqref{eq:kinematic_bicycle} converted to discrete time with a specified sampling period $\Delta T$ and using the Euler method.


\subsection{Reward function}\label{sec:prob_form_R}

The reward function $R(\bar{s}_t, \bar{u}_t)$ is a mathematical representation of the driving goals of the driver. Here, we start by considering the interactions between the ego vehicle and one other vehicle, i.e., $\bar{s}_t = (s_t^0, s_t^1)$ and $\bar{u}_t = (u_t^0, u_t^1)$. In this case, the traffic state is composed of the states of these two vehicles, and the reward received by the ego vehicle depends on the states and control inputs of both vehicles. Following \cite{liu2021cooperation}, we consider
\begin{equation} \label{eq:reward}
    R\big(\bar{s}_t, u_t^0, u_t^1\big) = w^T r,
\end{equation}
where $r = [r_1, r_2, r_3, r_4, r_5]^T$ and $w \in \mathbb{R}_+^5$ is a vector of weights. The reward terms $r_1,\dots,r_5$ are defined to represent the following common considerations during driving: 1) safety $(r_1, r_2)$, i.e., not colliding with other vehicles and not getting off the road; 2) liveness $(r_3, r_4)$, i.e., approaching the destination; and 3) comfort $(r_5)$, i.e., maintaining a reasonable separation from other vehicles. The reader is referred to \cite{liu2021cooperation} for more detailed definitions of $r_1,\dots,r_5$.

\begin{figure}[H]
\begin{center}
\begin{picture}(240.0, 105.0)
\put(  45,  -10){\epsfig{file=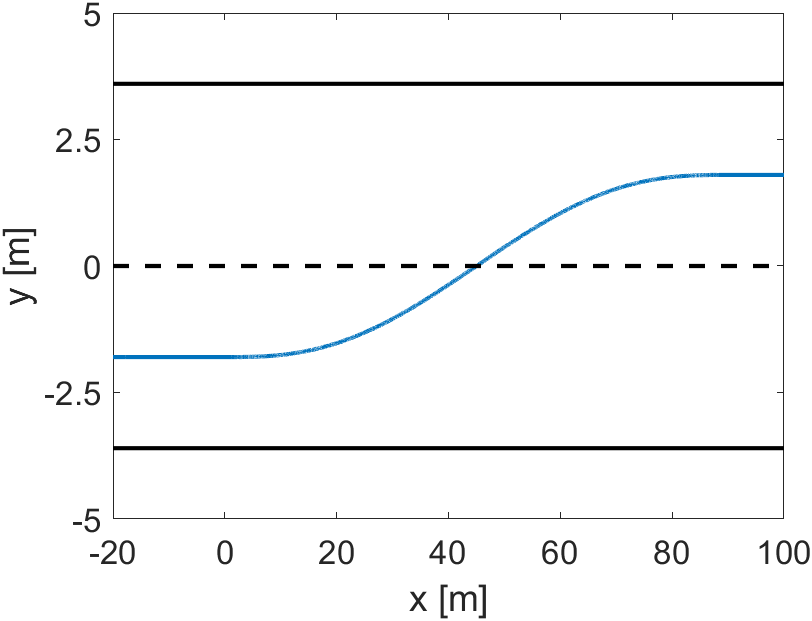,width=0.6\linewidth}}  
\end{picture}
\end{center}
      \caption{A sampled $5$th-order polynomial lane change trajectory.}
      \label{fig:polylanechange}
\end{figure}

\subsection{Selecting Trajectories as vehicle actions} \label{sec:gen_traj}


Instead of considering a discrete set of acceleration $a$ and steering $\delta_f$ levels as in \cite{liu2021cooperation}, we consider a sampled set of vehicle motion trajectories over a planning horizon of $T = N \Delta T$~[s] as the action space for each vehicle. Specifically, each trajectory is a time history of vehicle state $s_t = [x_t, y_t, v_t, \psi_t]^T$ starting from the vehicle's current state $s_0$. Note that the time history of control inputs $u_t = [a_t, \delta_{f,t}]^T$ corresponding to each trajectory can be calculated according to the vehicle dynamics model~\eqref{eq:kinematic_bicycle}. Compared to representing vehicle motion using discrete acceleration and steering levels as in \cite{liu2021cooperation}, the method here can lead to smoother trajectories and finer-resolution controls. 

For interacting vehicles driving in the target lane, we only consider their longitudinal motion, which corresponds to the assumption that these vehicles do not change lanes. Assuming $\psi = 0$ and $\delta_f = 0$, the kinematic bicycle model \eqref{eq:kinematic_bicycle} for these vehicles reduces to
\begin{equation}\label{eqn:double_int}
    \dot{x} = v, \quad \dot{y} = 0, \quad \dot{v} = a, \quad \dot{\psi} = 0, \quad \beta = 0.
\end{equation}
In this case, a trajectory starting with a given initial condition depends only on the profile of acceleration $a$ over $[0,T]$. In particular, at each sample time instant, we consider $81$ acceleration profiles, which translates into $81$ trajectories through \eqref{eqn:double_int}, for each interacting vehicle $k$ driving in the target lane, and we treat these trajectories as its admissible actions. Note that we also enforce the speed limits $v_t^k \in [v_{\min}, v_{\max}]$ when we generate these trajectories. We denote each of such trajectories as $\gamma_m^k(s_0^k)$, with $m = 1, 2,\dots, 81$, and the collection of such trajectories as $\Gamma^k(s_0^k):=\{\gamma_m^k(s_0^k)\}_{m=1}^{81}$.



The merging vehicle's maneuvers include both lane keeping and lane change. Trajectories or pieces of trajectories that represent lane keeping are generated using \eqref{eqn:double_int} in a similar way as above. For a lane change, we use $5$th-order polynomials to represent lane change trajectories \cite{papadimitriou2003fast}. Specifically, a lane change trajectory is produced by the solution to the following boundary value problem: Find the coefficients $a_1,\dots,a_5$ and $b_1,\dots,b_5$ such that the $5$th-order polynomials
\begin{equation} \label{eqn:5th_order_poly}
    \begin{aligned}
    x(\zeta) &= a_0 + a_1 \zeta + a_2 \zeta^2 + a_3 \zeta^3 + a_4 \zeta^4 + a_5 \zeta^5, \\
    y(\zeta) &= b_0 + b_1 \zeta + b_2 \zeta^2 + b_3 \zeta^3 + b_4 \zeta^4 + b_5 \zeta^5,
    \end{aligned}
\end{equation}
satisfy specified initial and terminal conditions $(x_\text{ini}, \dot{x}_\text{ini},$ $\ddot{x}_\text{ini}, y_\text{ini}, \dot{y}_\text{ini}, \ddot{y}_\text{ini})$ and $(x_\text{term}, \dot{x}_\text{term}, \ddot{x}_\text{term}, y_\text{term}, \dot{y}_\text{term}, \ddot{y}_\text{term})$, where $(x_\text{ini}, \dot{x}_\text{ini}, \ddot{x}_\text{ini}, y_\text{ini}, \dot{y}_\text{ini}, \ddot{y}_\text{ini})$ corresponds to either the vehicle's current state or its state at the start of a lane change, and $(x_\text{term}, \dot{x}_\text{term}, \ddot{x}_\text{term}, y_\text{term}, \dot{y}_\text{term}, \ddot{y}_\text{term})$ corresponds to the vehicle's state after the completion of a lane change. The variable $\zeta$ in \eqref{eqn:5th_order_poly} denotes continuous time. We let $\zeta = 0$ correspond to the current sample time instant and assume that 1) the vehicle can start a lane change at any sample time instant $\zeta =  t\, \Delta T$, with $t = 0,\dots,N-1$, over the planning horizon, and 2) a complete lane change takes a constant time duration of $T_{lc} = 3$~[s] \cite{papadimitriou2003fast}. Then, for the case where at the current sample time instant the vehicle is in the middle of a lane change (i.e., the vehicle started the lane change $\Delta T_{lc}$~[s] ago), $(x_\text{ini}, \dot{x}_\text{ini},$ $\ddot{x}_\text{ini}, y_\text{ini}, \dot{y}_\text{ini}, \ddot{y}_\text{ini})$ corresponds to the vehicle's current state and is satisfied by \eqref{eqn:5th_order_poly} at $\zeta = 0$, while $(x_\text{term}, \dot{x}_\text{term}, \ddot{x}_\text{term}, y_\text{term}, \dot{y}_\text{term}, \ddot{y}_\text{term})$ is satisfied by \eqref{eqn:5th_order_poly} at $\zeta = T_{lc} - \Delta T_{lc}$. For the case where the vehicle starts a lane change at a future sample time instant $t\, \Delta T$, $(x_\text{ini}, \dot{x}_\text{ini}, \ddot{x}_\text{ini}, y_\text{ini}, \dot{y}_\text{ini}, \ddot{y}_\text{ini})$ corresponds to the vehicle's state at the start of the lane change and is satisfied by \eqref{eqn:5th_order_poly} at $\zeta = t\, \Delta T$, while $(x_\text{term}, \dot{x}_\text{term}, \ddot{x}_\text{term}, y_\text{term}, \dot{y}_\text{term}, \ddot{y}_\text{term})$ is satisfied by \eqref{eqn:5th_order_poly} at $\zeta = t\, \Delta T + T_{lc}$. Furthermore, we allow the vehicle, when it is in the middle of a lane change, to abort the lane change at any sample time instant $\zeta =  t\, \Delta T$ over the planning horizon. This represents a ``change of mind'' of the driver when a previously planned lane change becomes no longer feasible/safe. A trajectory for aborting a lane change is generated in a similar way as a lane change trajectory, but the terminal condition $(x_\text{term}, \dot{x}_\text{term}, \ddot{x}_\text{term}, y_\text{term}, \dot{y}_\text{term}, \ddot{y}_\text{term})$ corresponds now to the vehicle's state after its returns to its original lane. Finally, we glue together pieces of trajectories for lane keeping, lane change, and aborting lane change to construct complete trajectories over the planning horizon. This way, we obtain a total of $162$ trajectories for the merging vehicle that we treat as admissible actions. Each of these trajectories is characterized by 1) whether and when to start a lane change and 2) whether and when to abort an improper lane change. Fig.~\ref{fig:polylanechange} illustrates a sampled set of such trajectories when the vehicle has not started a lane change and those when the vehicle is in the middle of a lane change. We denote each of such trajectories as $\gamma_m^0(s_0^0)$, with $m = 1, 2,\dots, 162$, and the collection of such trajectories as $\Gamma^0(s_0^0):=\{\gamma_m^0(s_0^0)\}_{m=1}^{162}$.


\begin{figure}[H]
\begin{center}
\begin{picture}(240.0, 95.0)
\put(  -10,  -10){\epsfig{file=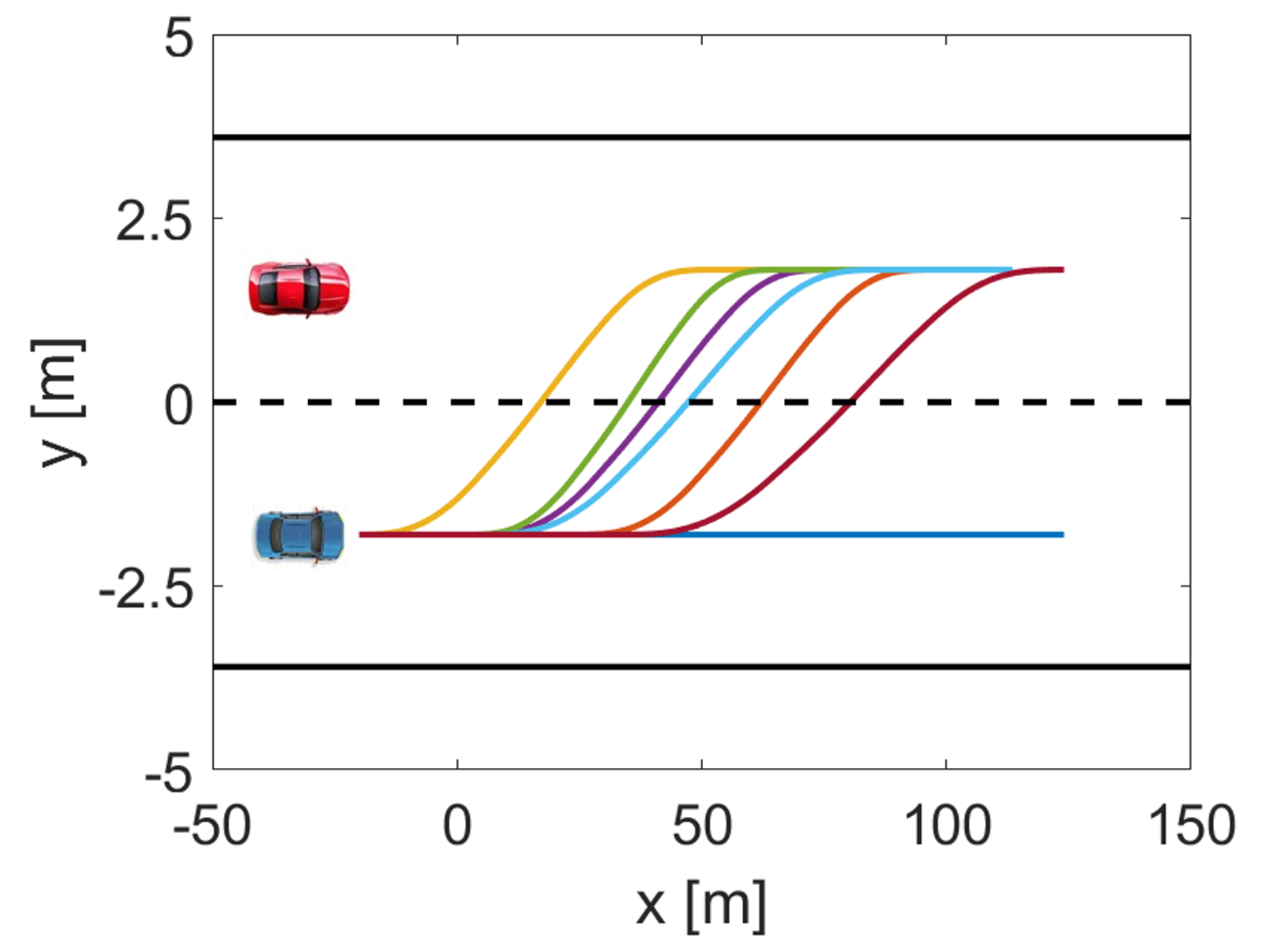,width=0.53\linewidth}}  
\put(  122,  -10){\epsfig{file=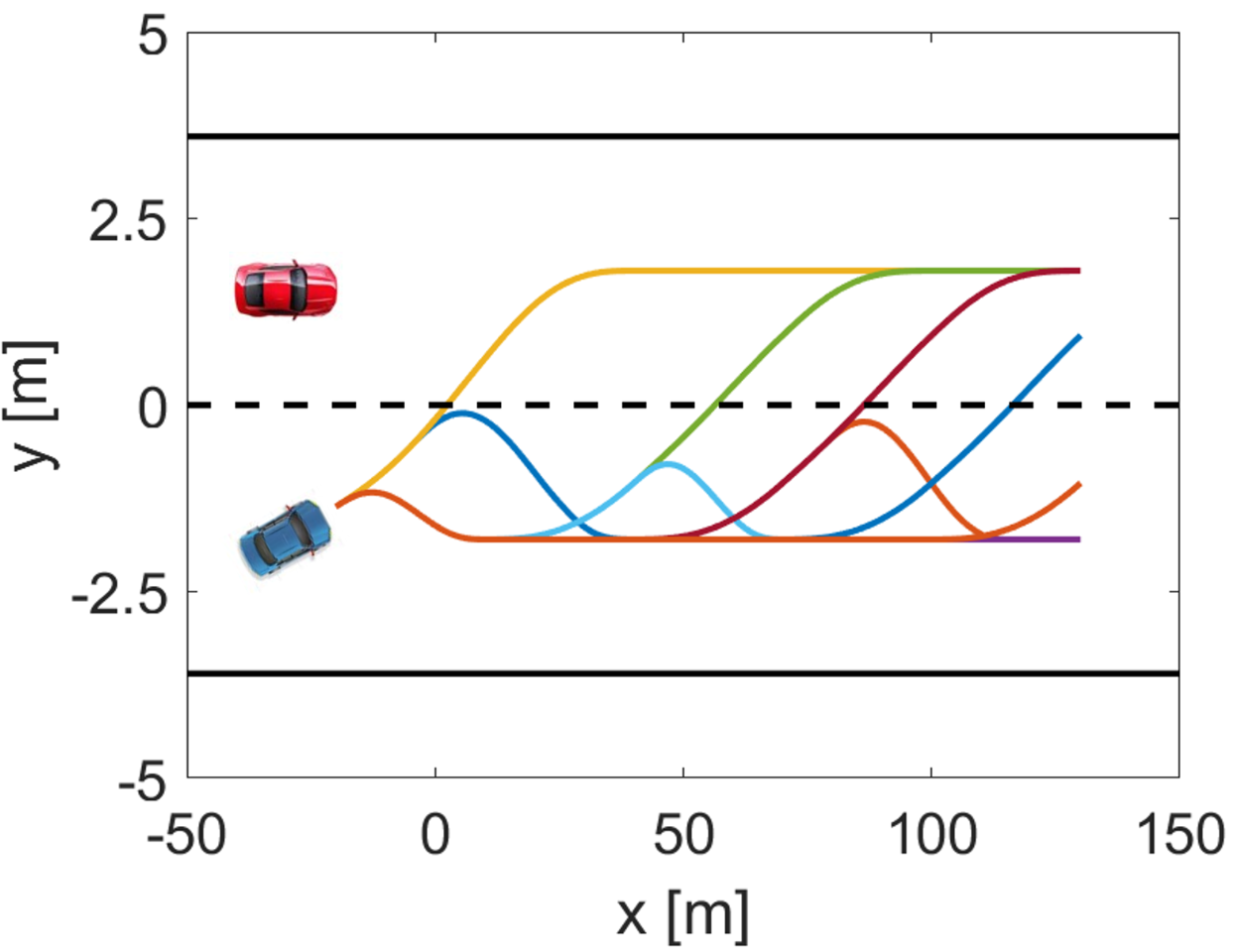,width=0.515\linewidth}}  
\end{picture}
\end{center}
      \caption{Sample trajectories for the merging vehicle.}
      \label{fig:polylanechange}
\end{figure}

We have defined the choice of a trajectory over a planning horizon for the merging vehicle as its action. Note that the time history of control inputs $u_t = [a_t, \delta_{f,t}]^T$ corresponding to each of these trajectories can be calculated according to the vehicle dynamics model~\eqref{eq:kinematic_bicycle}. In the actual implementation, the chosen trajectory can also be commanded to a lower level vehicle motion controller, and in this paper, it is assumed that the motion according to \eqref{eqn:5th_order_poly} is accurately realized.

\subsection{Model predictive control strategy}

We first consider an MPC-based trajectory planning strategy for the autonomous ego vehicle accounting for the presence of a signle interacting vehicle: At each sample time instant $t$, the ego vehicle computes an optimal trajectory, $(\gamma_t^0)^*$, that maximizes its cumulative reward over the planning horizon according to
\begin{align} 
    &(\gamma_t^0)^* \in \argmax_{\gamma_t^0 \in \Gamma^0(s^0_t)} \sum_{\tau = 0}^{N-1} \lambda^\tau R\big(\bar{s}_{t+\tau}, u^0_{t+\tau}, u^1_{t+\tau}\big), \label{eq:opt_reward}\\
    &\quad \text{s.t.} \quad \bar{s}_{t+\tau+1} = f\big(\bar{s}_{t+\tau}, u^0_{t+\tau}, u^1_{t+\tau}\big), \nonumber \\
    &\quad \quad\quad\, \bar{s}_{t+\tau} \in S_\text{safe}, \quad \forall \tau  = 1, \dots, N, \nonumber
\end{align}
where $\bar{s}_{t+\tau} = (s_{t+\tau}^0, s_{t+\tau}^1)$ represents the predicted traffic state at the discrete time instant $t + \tau$, while $u^0_{t+\tau}$ and $u^1_{t+\tau}$ represent, respectively, the predicted ego vehicle's and interacting vehicle's control inputs at $t + \tau$. The parameter $\lambda \in (0, 1)$ is discounting future reward thereby prioritizing immediate reward. In \eqref{eq:opt_reward}, $R\big(\bar{s}_{t+\tau}, u^0_{t+\tau}, u^1_{t+\tau}\big)$ represents the reward received by the ego vehicle at $t + \tau$, which is described in Section~\ref{sec:prob_form_R}, and $S_{\text{safe}}$ represents a set of {\it safe} traffic states, used to enforce strict safety specifications (such as collision avoidance, road boundary constraints, etc). After an optimal trajectory $(\gamma_t^0)^*$ is obtained, the ego vehicle applies the control inputs corresponding to this trajectory, $(u_t^0)^* = [(a_t^0)^*, (\delta_{f,t}^0)^*]^T$, over one sampling period to update its state, and then repeats the above procedure at the next sample time instant $t+1$.

Note the following points: 1) The expression \eqref{eq:opt_reward} corresponds to the case where the ego vehicle interacts with only one other vehicle ($k = 1$). We will extend our MPC strategy to handle multiple vehicle interactions in Section~\ref{sec:game_decision_making_M}. 2) The ego vehicle's control inputs over the planning horizon, $\{u^0_{t},\dots,u^0_{t+N-1}\}$, correspond to its planned trajectory $\gamma_t^0$ and are calculated using $\gamma_t^0$ and the vehicle dynamics model~\eqref{eq:kinematic_bicycle}, as has been discussed in Section~\ref{sec:gen_traj}. 3) The interacting vehicle's control inputs over the planning horizon, $\{u^1_{t},\dots,u^1_{t+N-1}\}$, are unknown variables. In what follows, we introduce a game-theoretic approach that enables predictions of $\{u^1_{t},\dots,u^1_{t+N-1}\}$ in response to the ego vehicle's actions so that make the MPC problem \eqref{eq:opt_reward} solvable.


%% file: sections/03_leader_follower_game.tex
\section{Game-Theoretic Model for Vehicle Cooperation Behaviors and Explicit Representation Using Imitation Learning} \label{sec:lf_game}

In this section, we introduce the leader-follower game employed in this paper for modeling the interaction/cooperation between the merging vehicle and vehicles driving in the target lane. In order to simplify the online computations associated with this game-theoretic model, imitation learning is utilized to derive a neural network-based explicit representation of the model, which is used online for predicting the interacting vehicles' trajectories in response to the merging ego vehicle's actions in our MPC-based trajectory planning strategy.

\subsection{Leader-follower game-theoretic model}

During a highway forced merge process, the merging vehicle (ego vehicle) interacts with other vehicles driving in the target lane, who may choose to proceed or yield to the merging vehicle depending on the traffic situation and individual driver's preference. In this paper, we consider a game-theoretic model based on pairwise leader-follower interactions, called a leader-follower game, to represent drivers' cooperation intentions and their resulting vehicle behaviors. In this model, a vehicle (or, a driver) who decides to proceed before another vehicle is a leader in this vehicle pair and the one who decides to yield to another vehicle is a follower in the pair. The leader and the follower use different decision strategies. This leader-follower game-theoretic model was originally proposed in \cite{nan2021gametheoretic}, where it demonstrated the ability to effectively model drivers' intentions to proceed or yield (e.g., caused by common traffic rules and etiquette) in driving through intersections scenarios. Here, we briefly review this game-theoretic model and introduce its application to our highway forced merge scenarios.



Denote the trajectories of the leader and the follower as $\gamma_{l,t} \in \Gamma_l(\bar{s}_t)$, and $\gamma_{f,t} \in \Gamma_f(\bar{s}_t)$, respectively, where $\Gamma_l(\bar{s}_t)$ and $\Gamma_f(\bar{s}_t)$ are the sets of admissible trajectories of the leader and the follower. We assume that both vehicles make decisions to maximize their cumulative rewards, denoted as $\mathbf{R}_l(\bar{s}_t, \gamma_{l,t}, \gamma_{f,t})$ and $\mathbf{R}_f(\bar{s}_t, \gamma_{l,t}, \gamma_{f,t})$, respectively, and defined according to
\begin{equation}
    \mathbf{R}_\sigma\big(\bar{s}_t, \gamma_{l,t}, \gamma_{f,t}\big) = \sum_{\tau = 0}^{N-1} \lambda^\tau R_\sigma\big(\bar{s}_{t+\tau}, u_{l,t+\tau}, u_{f,t+\tau} \big),
\end{equation}
where $\sigma \in L = \{\text{leader}, \text{follower}\}$ represents the leader or follower role in the game, $R_\sigma\big(\bar{s}_{t+\tau}, u_{l,t+\tau}, u_{f,t+\tau} \big)$ is the reward function for the leader or the follower defined as in Section~\ref{sec:prob_form_R}, and $u_{l,t+\tau}$ and $u_{f,t+\tau}$, $\tau = 0,\dots,N-1$, are the control inputs corresponding to $\gamma_{l,t}$ and $\gamma_{f,t}$ as described in Section~\ref{sec:gen_traj}.



Specifically, we model the leader's and the follower's interactive decision processes as follows:
\begin{align}
    &\gamma_l^*(\bar{s}_t) \in \argmax_{\gamma_{l,t} \in \Gamma_l(\bar{s}_t)}\, Q_l(\bar{s}_t, \gamma_{l,t}), \label{eq:lf_leader_opt_act} \\
    &\gamma_f^*(\bar{s}_t) \in \argmax_{\gamma_{f,t} \in \Gamma_f(\bar{s}_t)}\, Q_f(\bar{s}_t, \gamma_{f,t}), \label{eq:lf_follower_opt_act}
\end{align}
where $\gamma_l^*(\bar{s}_t)$ (resp. $\gamma_f^*(\bar{s}_t)$) is an optimal trajectory of the leader (resp. follower) given the current traffic state $\bar{s}_t$, and $Q_l$ and $Q_f$ are defined as
\begin{align}
    &Q_l(\bar{s}_t, \gamma_{l,t}) = \min_{\gamma_{f,t} \in \Gamma_f^*(\bar{s}_t)} \mathbf{R}_l(\bar{s}_t, \gamma_{l,t},\gamma_{f,t}), \label{eq:lf_Q_leader}\\
    &Q_f(\bar{s}_t, \gamma_{f,t}) = \min_{\gamma_{l,t} \in \Gamma_l(\bar{s}_t)} \mathbf{R}_f(\bar{s}_t, \gamma_{l,t},\gamma_{f,t}), \label{eq:lf_Q_follower}
\end{align}
where $\Gamma_f^*(\bar{s}_t) = \{ \gamma_{f,t}' \in \Gamma_f(\bar{s}_t): Q_f(\bar{s}_t, \gamma_{f,t}') \geq Q_f(\bar{s}_t,\gamma_{f,t}),$ $\forall \gamma_{f,t} \in \Gamma_f(\bar{s}_t) \}$.

The decision model \eqref{eq:lf_leader_opt_act}-\eqref{eq:lf_Q_follower} can be explained as follows: A follower represents a driver who intends to yield. Due to uncertainty about the other driver's action, the follower decides to take an action that maximizes her worst-case reward through \eqref{eq:lf_follower_opt_act} and \eqref{eq:lf_Q_follower}. Such a ``max-min'' decision strategy of the follower models the yielding behavior because it assumes the other driver can take actions freely. Similarly, a leader represents a driver who intends to proceed and assumes the other driver will yield. Therefore, the leader uses the follower model to predict the other driver's action and takes an action that maximizes the leader own reward under the predicted follower's action through \eqref{eq:lf_leader_opt_act} and \eqref{eq:lf_Q_leader}. This leader-follower game model is partly inspired by the Stackelberg game model \cite{bacsar1998dynamic}, but relaxes several assumptions of the Stackelberg model that generally do not hold for driver interactions in traffic. The reader is referred to \cite{nan2021gametheoretic} for more discussions of this leader-follower game model and of its effectiveness for modeling driver interactions in multi-vehicle scenarios.

Note that although the asymmetric leader-follower roles in the decision model \eqref{eq:lf_leader_opt_act}-\eqref{eq:lf_Q_follower} are used to represent drivers' intentions to proceed and yield, respectively, the model does not imply that a leader interacting vehicle will always force a merging vehicle to merge behind it or a follower interacting vehicle will always let a merging vehicle merge in front of it. For instance, a merging vehicle may merge in front of a leader interacting vehicle in the following two situations: 1) The merging vehicle is ahead of the interacting vehicle with a sufficiently large distance to allow safe merging. 2) The merging vehicle is about to reach the end of its lane. Because getting off the road yields a large penalty (see Section~\ref{sec:prob_form_R}), the merging vehicle may choose to merge ahead of the interacting vehicle to avoid the large penalty as long as its merging will not lead to a collision between the two vehicles (it will not merge if the merging will cause a collision because the penalty for collision is even larger than for getting off the road). The above observations clarify that the leader-follower roles in our decision model \eqref{eq:lf_leader_opt_act}-\eqref{eq:lf_Q_follower} are not assigned by vehicle spatial positions (i.e., a leader is not necessarily a vehicle in front). Moreover, this model allows a merging vehicle to force the traffic in the target lane to let it merge into: As the merging vehicle approaches the end of its lane, it is increasingly inclined to merge to avoid the penalty for getting off the road even if all of the interacting vehicles in the target lane are leaders (i.e., their drivers all originally intend to proceed) and the current gaps are not large enough in terms of comfort. The model \eqref{eq:lf_leader_opt_act}-\eqref{eq:lf_Q_follower} enables these leader interacting vehicles to predict the merging vehicle's upcoming merging maneuver. Then, for their own safety and comfort, they will slow down to enlarge the gap and, consequently, warrant the merging. Therefore, our leader-follower model \eqref{eq:lf_leader_opt_act}-\eqref{eq:lf_Q_follower} is suitable for trajectory prediction and planning in forced merge scenarios.

\subsection{Explicit representation of leader-follower game policy through imitation learning}

Based on \eqref{eq:lf_leader_opt_act}-\eqref{eq:lf_Q_follower}, we are able to predict other vehicles' decision and trajectories given the knowledge of drivers' intentions and the current traffic state information. Hence, we can denote leader's optimal action policy as $\gamma_l^*(\bar{s}_t)$ and follower's optimal action policy as $\gamma_f^*(\bar{s}_t)$. Obtaining $\gamma_l^*(\bar{s}_t)$ and $\gamma_f^*(\bar{s}_t)$ require going through \eqref{eq:lf_leader_opt_act}-\eqref{eq:lf_Q_follower}, and the repeated online computations involving \eqref{eq:lf_leader_opt_act}-\eqref{eq:lf_Q_follower} can be time consuming. As a result, we want to explicitly represent $\gamma_l^*$ and $\gamma_f^*$.

Here, $\gamma_\sigma^*(\bar{s}_t), \sigma \in L$ are maps that map current traffic state to a predicted trajectory that other vehicle will follow. These maps are determined according to \eqref{eq:lf_leader_opt_act}-\eqref{eq:lf_Q_follower}. Instead of algorithmically determining $\gamma_l^*(\bar{s}_t)$ and $\gamma_f^*(\bar{s}_t)$, we follow \cite{tian2020game} and exploit the use of supervised learning, more specifically, imitation learning, to represent $\gamma_\sigma^*(\bar{s}_t)$.

Imitation learning can be considered as a supervised learning problem, where an autonomous agent tried to learn a policy by observing expert's demonstrations. The expert demonstration can be generated either by a human operator or an artificial intelligent agent. In this work, we treat $\gamma_\sigma^*(\bar{s}_t)$ obtained by \eqref{eq:lf_leader_opt_act}-\eqref{eq:lf_Q_follower} as the expert policy. 

In particular, the ``Dataset Aggregation'' algorithm \cite{ross2011reduction} has been utilized to obtain an imitated policy $\hat{\gamma}_\sigma$. The overall learning objective for the Dataset Aggregation algorithm can be described by,
\begin{align}
    &\hat{\gamma}_\sigma \in \argmin_{\gamma_\theta} \mathbb{E}_{\bar{s} \sim \mathbb{P}(\bar{s}|\hat{\gamma}_\theta)} \Big(\mathcal{L}(\gamma^*_\sigma(\bar{s}), \gamma_\theta(\bar{s}))\Big), \label{eq:dataset_aggregation}\\
    &\mathbb{E}_{\bar{s} \sim \mathbb{P}(\bar{s}|\gamma_\theta)}(\cdot) = \int (\cdot)\, \text{d}\mathbb{P}(\bar{s} |\gamma_\theta).
\end{align}
where $\gamma_\theta$ represents a policy with respect to which optimization is performed and which is parameterized by $\theta$ (e.g. neural network weights), and $\mathcal{L}$ represents a loss function. More detailed discussions on the imitation learning and the ``Dataset Aggregation'' algorithm can be found in \cite{tian2020game} and \cite{ross2011reduction}.

The model \eqref{eq:lf_leader_opt_act}-\eqref{eq:lf_Q_follower} and the imitation learning policies can be used to predict the other vehicles' decisions and future trajectories under the knowledge of their drivers' cooperation intentions. However, in a given traffic scenario, we may not know the other drivers' cooperation intentions a priori, because a driver's intention depends not only
on the traffic situation (e.g., the relative position and velocity between two vehicles) but also on the driver's style/type (e.g., aggressive versus conservative). To deal with prior uncertainties about other vehicles' cooperation intentions, in what follows we describe an approach where
such uncertainties are modeled as latent variables and the autonomous vehicle planning and control problem exploits estimating other vehicle's cooperation intentions as well as using predictive control method to obtain the optimal trajectory. 

%% file: sections/04_decision_making_algorithm.tex
\section{Decision Making under Cooperation Intention Uncertainty} \label{sec:game_decision_making}

In this section, we describe the decision making algorithm, called the Leader-Follower Game Controller (LFGC), for the highway forced merge scenario under cooperation intention uncertainty. During the forced merge process, we generate an estimate of other driver's cooperation intention, as described in this section. Based on the estimate of cooperation intention, we apply the control strategy presented in \eqref{eq:opt_reward} under multi-vehicle interactions settings by considering pairwise interactions.

\subsection{Estimation of interacting vehicle's cooperation intention}

According to Section~\ref{sec:lf_game}, we can model other driver's behavior based on their cooperation intentions using the leader-follower game. A yielding vehicle may have similar behavior as a follower in the game, while a proceeding (not yielding) vehicle may be modeled as a leader in the game. In this sense, we can estimate interacting vehicle's cooperation intention by estimating their leader or follower roles in the leader-follower game.

To achieve that, we consider the traffic dynamics model \eqref{eq:sys_dyn} and the leader or follower's optimal actions \eqref{eq:lf_leader_opt_act} and \eqref{eq:lf_follower_opt_act}. From the perspective of the ego vehicle, the interacting vehicle is playing a leader-follower game with it, and the traffic dynamics model can be written as
\begin{equation} \label{eq:sys_dyn_lf}
    \bar{s}_{t+1} = f(\bar{s}_t, u_t^0, u_t^1) = f\big(\bar{s}_t, u_t^0, (u_{\sigma,t}^{1})^*(\bar{s}_t) \big),
\end{equation}
where $u_t^0$ is the control of ego vehicle, $u_t^1$ is the control of the interacting vehicle and is determined by the leader-follower game, $\sigma \in L =  \{\text{leader}, \text{follower}\}$ represents either leader of follower, and $(u_{\sigma,t}^{1})^*(\bar{s}_t)$ is the first control input corresponding to the optimal trajectory of $\gamma_\sigma^*(\bar{s}_t)$ in \eqref{eq:lf_leader_opt_act} and \eqref{eq:lf_follower_opt_act}. Now the only input to \eqref{eq:sys_dyn_lf} is the control of the ego vehicle $u_t^0$.

However, in reality, the interacting vehicle's decision does not necessarily follow the optimal policy computed from \eqref{eq:lf_leader_opt_act} and \eqref{eq:lf_follower_opt_act}. In order to account for the difference between the leader-follower policy and the actual policy of the interacting vehicle, we assume the system is propagated by \eqref{eq:sys_dyn_lf} with an additive Gaussian noise, i.e.,
\begin{equation} \label{eq:sys_dyn_lf_noise}
    \bar{s}_{t+1} = f\big(\bar{s}_t, u_t^0, (u_{\sigma,t}^{1})^*(\bar{s}_t) \big) + w, \quad w \sim N(0, W),
\end{equation}
where $w$ is the additive Gaussian noise with 0 mean and covariance $W$.

The ego vehicle is assumed to have a prior belief on $\sigma$, denoted as $\mathbb{P}(\sigma = l)$, with $l \in L =  \{\text{leader}, \text{follower}\}$. Then based on all previous traffic states and on all actions taken by the ego vehicle, 
\begin{equation}
    \xi_t = \{\bar{s}_0, \bar{s}_1, \dots, \bar{s}_t, u_0^0, u_1^0, \dots, u_{t-1}^0 \},
\end{equation}
the ego vehicle needs to compute or maintain a posterior belief of interacting vehicle's leader or follower role, $\mathbb{P}(\sigma = l | \xi_t)$.

The conditional posterior belief of interacting vehicle's leader or follower's role is computed using the hybrid estimation algorithm proposed in \cite{hwang2006state}. 

Specifically, identification of the interacting vehicle's leader or follower role can be achieved by,
\begin{equation} \label{eq:belief_update_orignal}
\mathbb{P}(\sigma = l| \xi_t) = \frac{\Lambda_{l,t} }{c_t} \sum_{k\in L}\pi_{lk} \mathbb{P}(\sigma = k|\xi_{t-1}),      
\end{equation}
where $\mathbb{P}(\cdot|\cdot)$ is the conditional probability; $\pi_{lk}$ denotes the transition probability of the interaction vehicle's role from $k$ to $l$; and $\Lambda_{l,t}$ is the likelihood function of the interacting vehicle as role $l$, defined by,
\begin{equation} 
\begin{aligned}
    &\Lambda_{l,t} = \mathcal{N}(r_{l,t}, 0, W), \\
    &r_{l,t} = \bar{s}_t - f(\bar{s}_{t-1}, u_{t-1}^0, (u_{l,t}^{1})^*(\bar{s}_{t-1})),
\end{aligned}
\end{equation}
where $\mathcal{N}(r_{l,t}, 0, W)$ denotes the probability density function of the normal distribution with mean 0 and covariance $W$ evaluated at $r_{l,t}$; and $c_t$ is the normalization constant. 

Assuming the interacting vehicle's role remains unchanged over the merge period, i.e., $\pi_{lk} = 1$ when $l = k$ and $\pi_{lk} = 0$ when $l \neq k$, the posterior belief of the interacting vehicle's leader or follower role can be updated using the following equation,
\begin{equation} \label{eq:belief_update}
    \mathbb{P}(\sigma = l | \xi_t) =  \frac{\mathcal{N}(r_{l,t}, 0, W)\mathbb{P}(\sigma = l | \xi_{t-1})}{\sum_{k\in L} \mathcal{N}(r_{k,t}, 0, W) \mathbb{P}(\sigma = k | \xi_{t-1})}
\end{equation}
where $\mathbb{P}(\sigma = l | \xi_{t-1})$ is the previous belief of the interacting vehicle's leader or follower role.

\subsection{Control strategy for multi-vehicle interactions}\label{sec:game_decision_making_M}

When the traffic is busy, there may exist multiple vehicles on highway that may interfere with the ego vehicle's merge, such as in the case shown in Fig.~\ref{fig:intro_merge}. One low complexity solution would be for the ego vehicle to only consider interaction with the first vehicle, and after the first vehicle becomes farther away, the ego vehicle starts to interact with the second vehicle. However, this may slow down the estimate of later vehicles' intentions, and in the case of highway forced merge, this can lead to the ego vehicle missing an opportunity to merge. 

Another solution is to interact with multiple vehicles at the same time. In this case, a model needs to be constructed to predict interacting vehicle's actions. Although 2-player leader-follower game described in Section~\ref{sec:lf_game} can be extended to multi-player leader-follower game by considering a multi-level decision hierarchy and then solving for the Nash-equilibrium, such extensions may exponentially increase the computational time as the number of players increase. The Stackelberg equilibrium can be hard to obtain when there are more than 3 players \cite{yoo2018predictive}. As a result, we propose a computationally tractable approach to extend the framework to multi-vehicle interactions by considering pairwise interactions.

When there are $m$ interacting vehicles, we consider pairwise interactions of the ego vehicle and each interacting vehicle. Then we can construct $m$ traffic states denoted as $\bar{s}^k, \; k \in \{1, \dots, m\}$ which contains the states of the ego vehicle and $k$th interacting vehicle, and the dynamic model of each $\bar{s}^k$ is given by
\begin{equation}
    \bar{s}^k_{t+1} = f(\bar{s}^k_t, \bar{u}^k_t) = f(\bar{s}^k_t, u^0_t, u^k_t).
\end{equation}

Similarly, we can denote by $\sigma^k \in L =  \{\text{leader}, \text{follower}\}$ the pairwise leader or follower role of the $k$th interacting vehicle and by $\xi_t^k$ the collection of all previous pairwise traffic states and actions taken by the ego vehicle, i.e.,
\begin{equation}
    \xi_t^k = \{\bar{s}^k_0, \bar{s}^k_1, \dots, \bar{s}^k_t, u_0^0, u_1^0, \dots, u_{t-1}^0 \}.
\end{equation}

Then we can utilize \eqref{eq:belief_update} to update the belief of each interacting vehicle's leader or follower role, $\mathbb{P}(\sigma^k = l | \xi_t^k), l \in L =  \{\text{leader}, \text{follower}\}$. The MPC-based control strategy presented in \eqref{eq:opt_reward} can be reformulated as
\begin{align} 
    &(\gamma^0)^* \!\in\! \argmax_{\substack{\gamma^0 \in \Gamma^0(s^0_t)}} \sum_{k=1}^m \mathbb{E}\! \bigg\{\sum_{\tau = 0}^{N-1}\! \lambda^\tau R\big(\bar{s}^k_{t+\tau}, u^0_{t+\tau}, \hat{u}_{\sigma^k,t}^{k} (\bar{s}_{t+\tau}^k)\big) \bigg| \xi^k_t \bigg\}, \label{eq:opt_expect_reward} \\
    &\quad \text{s.t.} \quad \bar{s}_{t+\tau+1}^k = f\big(\bar{s}_{t+\tau}^k, u_{t+\tau}^0, \hat{u}_{\sigma^k,t}^{k} (\bar{s}_{t+\tau}^k)\big), \forall k = 1, \dots, m \nonumber \\
    &\quad \quad\quad\, \sum_{k=1}^m \mathbb{P} \Big(\bar{s}_{t+\tau}^k \in S_\text{safe}, \forall \tau  = 1, \dots, N\, \Big| \xi^k_t \Big) \geq m - \varepsilon, \nonumber 
\end{align}
where $\hat{u}_{\sigma,t}^{k}(\bar{s}_{t+\tau}^k)$ is the first control input corresponding to the trajectory of the trained policy $\hat{\gamma}_\sigma(\bar{s}_{t+\tau}^k)$ in \eqref{eq:dataset_aggregation}, and $\varepsilon \in [0, 1]$ represents a (user-specified) required probability level of constraint satisfaction.

The expectation in the objective function can be computed according to
\begin{equation}
\begin{aligned}
    &\mathbb{E}\bigg\{\sum_{\tau = 0}^{N-1}\! \lambda^\tau R\big(\bar{s}^k_{t+\tau}, u^0_{t+\tau}, \hat{u}_{\sigma^k,t}^{k} (\bar{s}_{t+\tau}^k)\big) \bigg| \xi^k_t \bigg\} \\ 
    &= \sum_{l \in L} \sum_{\tau = 0}^{N-1}\! \lambda^\tau R\big(\bar{s}^k_{l, t+\tau}, u^0_{t+\tau}, \hat{u}_{l,t}^{k} (\bar{s}_{l,t+\tau}^k)\big) \mathbb{P}(\sigma^k = l | \xi^k_t),
\end{aligned}
\end{equation}
where $\bar{s}_{l,t+\tau}^k$ is the predicted traffic state given that the interacting vehicle's role is $l$,
\begin{equation}
    \bar{s}_{l, t+\tau+1}^k = f\big(\bar{s}_{l, t+\tau}^k, u_{t+\tau}^0, \hat{u}_{l,t}^{k} (\bar{s}_{l, t+\tau}^k)\big)
\end{equation}
and the last constraint in \eqref{eq:opt_expect_reward} can be evaluated by,
\begin{equation}
\begin{aligned}
    &\mathbb{P} \Big(\bar{s}_{t+\tau}^k \in S_\text{safe}, \forall \tau  = 1, \dots, N\, \Big| \xi^k_t \Big) \\
    &= \sum_{l\in L} \min_{1 \leq \tau\leq N} \mathbb{I}_{S_\text{safe}}(\bar{s}_{l, t+\tau}^k ) \mathbb{P}(\sigma^k = l | \xi^k_t),
\end{aligned}
\end{equation}
where $\mathbb{I}_B(b)$ is the indicator function of $b$ in set $B$.

Note that the last constraint in \eqref{eq:opt_expect_reward} enforces the following condition,
\begin{equation}\label{eq:multi_veh_const}
    \mathbb{P}\Big(\bigcap_{k = 1}^m  \bar{s}^k_{t+\tau} \in S_\text{safe}, \forall \tau = 1, \dots, N\, \Big| \xi^k_t \Big) \geq 1 - \varepsilon,
\end{equation}
which means that the probability of any pairwise interactions entering unsafe states (e.g., collision and out of road boundaries) is less than $\varepsilon$. 

To derive \eqref{eq:multi_veh_const}, we first denote the event $A^k := \{ \bar{s}_{t+\tau}^k \in S_\text{safe}, \forall \tau = 1, \dots, N | \xi^k_t \}$, then 
\begin{align*}
    \mathbb{P}\big(\bigcap_{k=1}^m A^k \big) &= 1 - \mathbb{P}\big(\bigcup_{k=1}^m (A^k)^c \big) 
    \geq 1 - \sum_{k=1}^m \mathbb{P}\big( (A^k)^c \big) \\
    &= 1 - \sum_{k=1}^m ( 1 - \mathbb{P}(A^k) ) = \sum_{k=1}^m \mathbb{P}(A^k) - m + 1, 
\end{align*}
and applying the last constraint in \eqref{eq:opt_expect_reward}, it follows that
\begin{align*}
    \mathbb{P}\big(\bigcap_{k=1}^m A^k \big) \geq  \sum_{k=1}^m \mathbb{P}(A^k) - m + 1 \geq 1-\varepsilon.
\end{align*}

The major differences between \eqref{eq:opt_reward} and \eqref{eq:opt_expect_reward} are the following: 1) $\{u_t^1, u^1_{t+1}, \dots, u^1_{t+N-1}\}$ presented in \eqref{eq:opt_reward} are unknown, while in \eqref{eq:opt_expect_reward}, they are obtained based on trained policy from the imitation learning; 2) The maximization of the cumulative reward in \eqref{eq:opt_reward} is changed to maximization of the expected cumulative reward in \eqref{eq:opt_expect_reward} to account for probabilistic belief about the interacting vehicle's leader/follower role; 3) The expected cumulative reward is changed to the sum of the expected reward of all pairwise interactions to account for uncertain behavior of multiple vehicles; 4) The hard constraint is changed to a probabilistic chance constraint with $\varepsilon \in [0, 1]$ being a design parameter.

The decision making algorithm proceeds as follows: At the sampling time $t$, the ego vehicle measures the current states of each pairwise interaction and adds them together with the previous control input to the observation vectors $\xi_t^k$. The belief about each vehicle's leader or follower role is updated according to \eqref{eq:belief_update} based on $\xi_t^k$. Then, the MPC-based control strategy \eqref{eq:opt_expect_reward} is utilized to obtain the optimal trajectory $(\gamma^0)^*$ by searching through all trajectories introduced in Section~\ref{sec:gen_traj}, and the ego vehicle applies the first control input $(u_t^0)^*$ over one sampling period to update its states. The whole procedure is repeated at the next sampling time.


Note that the control strategy \eqref{eq:opt_expect_reward} is ``interaction-aware'' due to the following reasons: 1) It predicts other vehicles' trajectories under varied interaction intentions based on the leader-follower game-theoretic model \eqref{eq:lf_leader_opt_act}-\eqref{eq:lf_Q_follower}. 2) The predictions are closed-loop. Specifically, for different trajectory plans of the ego vehicle, $\gamma^0 \in \Gamma^0(s^0_t)$, the corresponding trajectory predictions of the other vehicles under certain intentions are different. This is the case because the predicted other vehicles' actions are traffic state-dependent while the predicted traffic states depend on the planned ego vehicle's trajectory. 3) The objective function in \eqref{eq:opt_expect_reward} is a conditional expectation and the constraint to represent safety is a conditional probability, both of which are conditioned on the latest estimates of other vehicles' intentions (i.e., leader or follower), $\mathbb{P}(\sigma^k = l | \xi^k_t)$. Meanwhile, the other vehicles' intentions are estimated based on their previous interaction behaviors.

%% file: sections/05_results.tex
\section{Simulation and Validation Results} \label{sec:results}

In this section, we present validation results of applying the proposed Leader-Follower Game Controller (LFGC) for autonomous vehicle forced merge problems. Specifically, we consider three simulation validations, and in these simulations, the LFGC assumes that interacting vehicles are playing leader-follower game with the ego vehicle and estimates their leader/follower roles in the game. We also assume that once in the mandatory lane change situation, the ego vehicle prepositions itself towards the lane marker with turn signals to declare its merge intention and starts the forced merge process. As a result, interacting vehicles are aware of the ego vehicle's merging intention and react accordingly. We first validate the LFGC with interacting vehicles controlled by leaders or followers in the leader-follower game. Then we test the LFGC versus interacting vehicles controlled by other types of drivers or actual traffic data. Specifically, we test the cases where interacting vehicles are controlled by intelligent driver model (IDM) and where interacting vehicles are following the actual US Highway 101 traffic data present in the Next Generation Simulation website \cite{us101data}. Note that our simulations are performed in MATLAB R2019a on an PC with Intel Xeon E3-1246 v3 @ 3.50 GHz CPU and 16 GB RAM.

\subsection{Interacting vehicles driven by leader/follower} \label{sec:res_lf}

\begin{figure}[h]
\begin{center}
\begin{picture}(240.0, 50.0)
\put(  -5,  -10){\epsfig{file=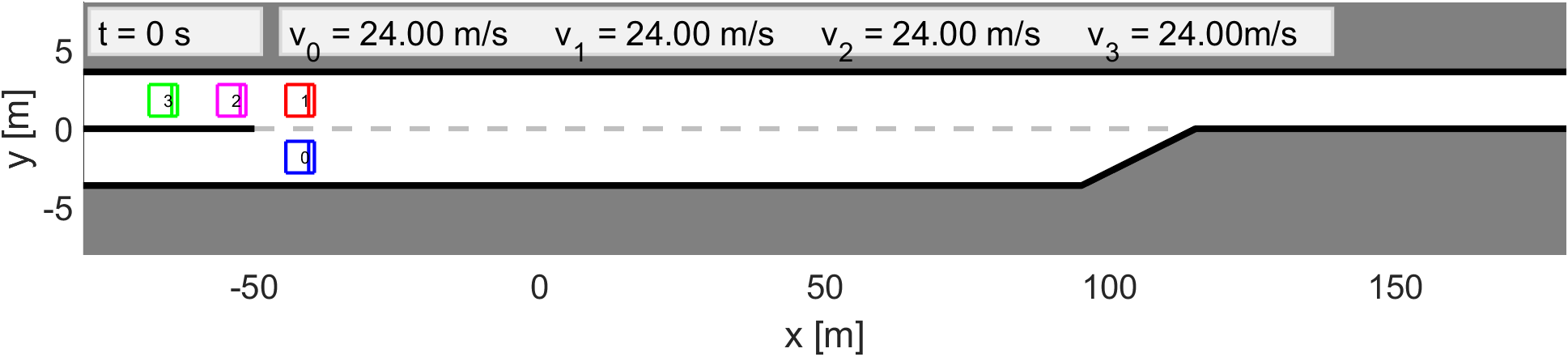,width=1.00\linewidth}}  
\end{picture}
\end{center}
      \caption{Illustration of highway forced merge scenarios for validations of the LFGC when highway vehicles are controlled by leaders/followers in the game.}
      \label{fig:scenario_LF}
      \vspace{-0.1in}
\end{figure}

We first test our proposed LFGC when interacting vehicles are simulated and controlled by leaders/followers in the game. The scenario we considered is shown in Fig.~\ref{fig:scenario_LF}, where the autonomous ego vehicle (blue) in the acceleration lane needs to merge onto the highway before the end of acceleration lane while multiple other vehicles (red, pink, green) are currently driving on the highway. The ego vehicle starts the forced merge process by biasing towards lane markers and flashing turn signals at the moment shown in Fig.~\ref{fig:scenario_LF}. In such a scenario, the autonomous vehicle needs to interact with other vehicles to merge safely.

\begin{figure*}[t]
\begin{center}
\begin{picture}(480.0, 360.0)
\put(  -5,  260){\epsfig{file=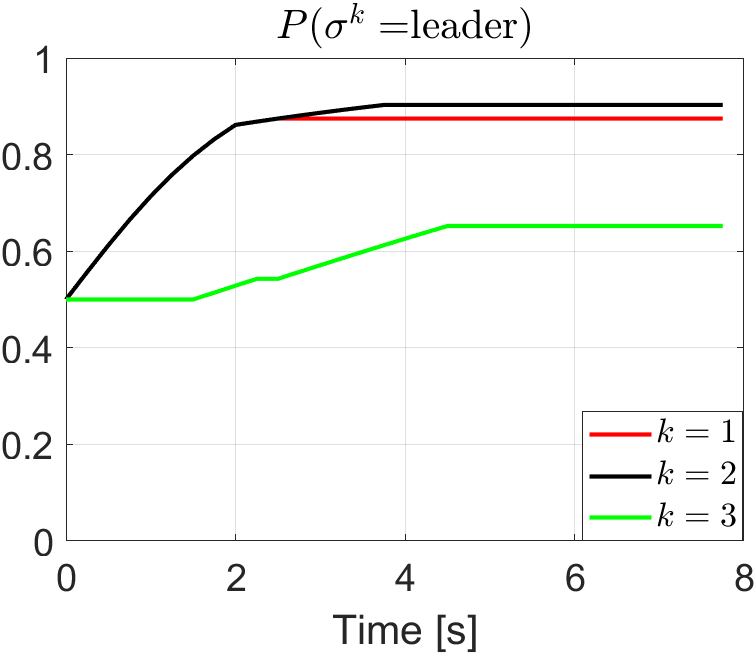,width=0.2\linewidth}}  
\put(  -5, 345){\text{(a-1)}}
\put(  110,  260){\epsfig{file=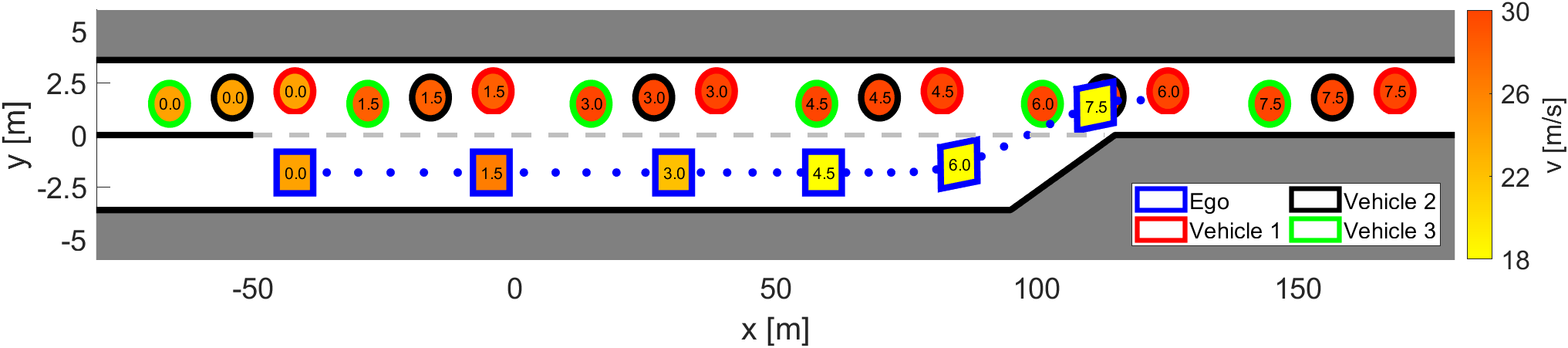,width=0.75\linewidth}}  
\put(  110, 345){\text{(a-2)}}
\put(  -5,  170){\epsfig{file=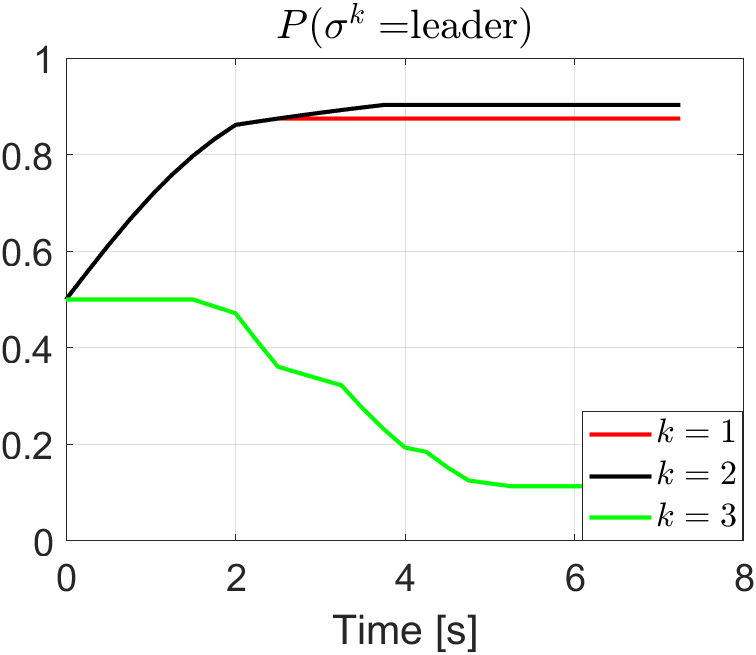,width=0.2\linewidth}}  
\put(  -5, 255){\text{(b-1)}}
\put(  110,  170){\epsfig{file=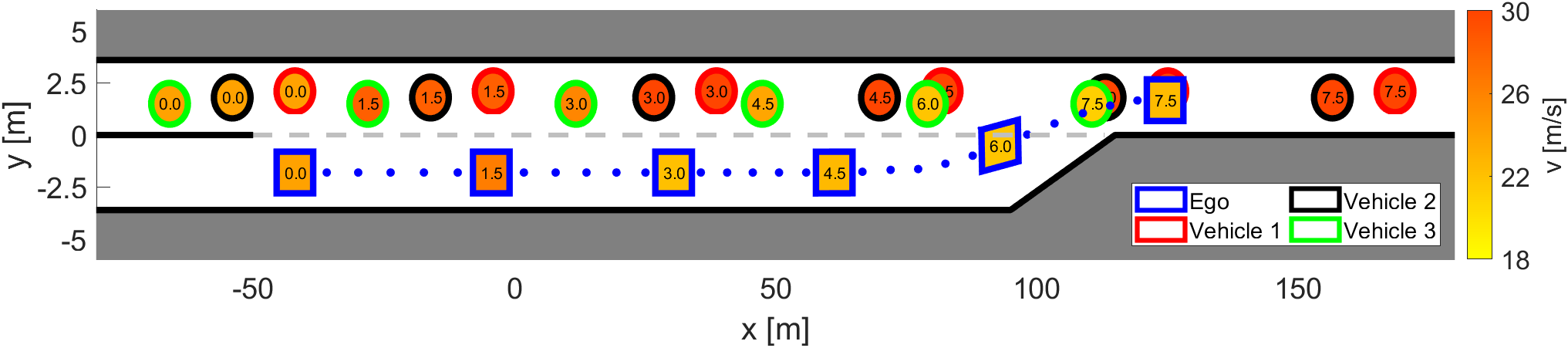,width=0.75\linewidth}}  
\put(  110, 255){\text{(b-2)}}
\put(  -5,  80){\epsfig{file=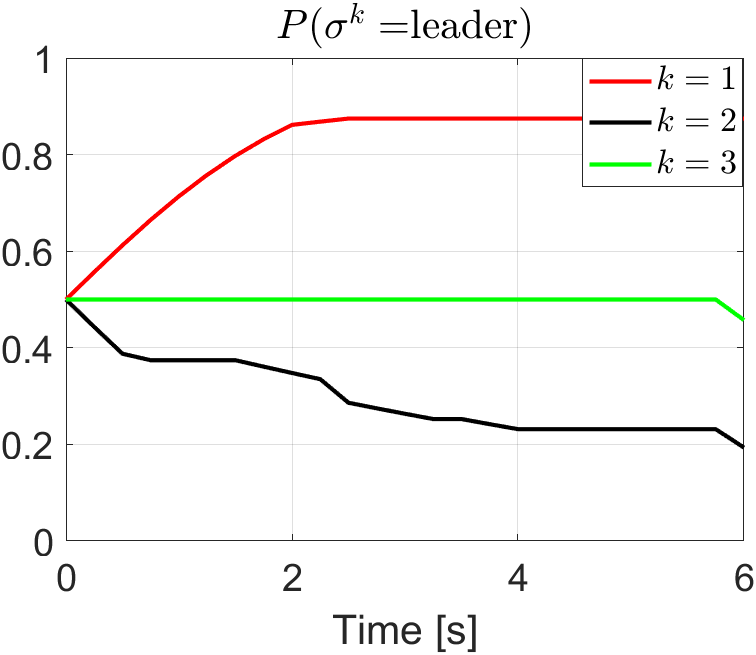,width=0.2\linewidth}}  
\put(  -5, 165){\text{(c-1)}}
\put(  110,  80){\epsfig{file=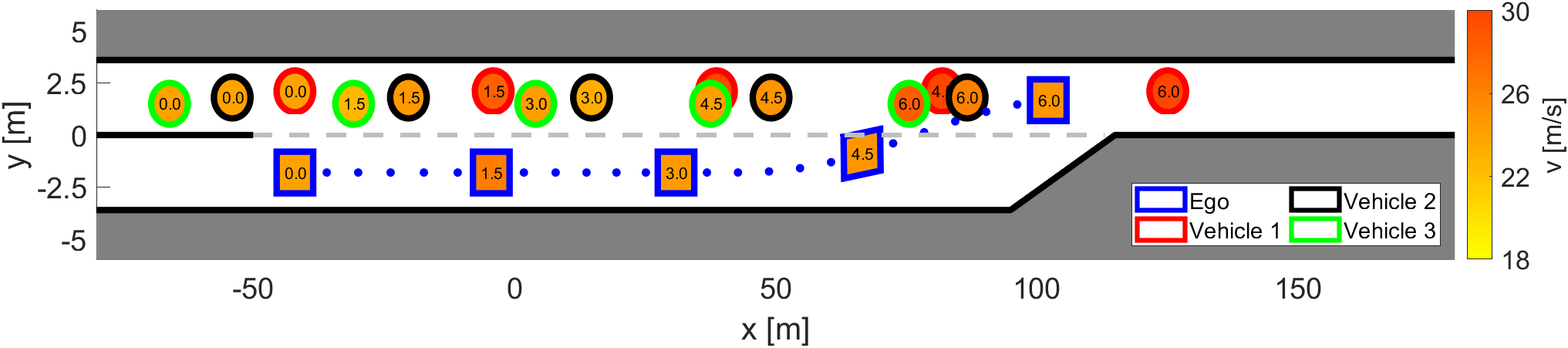,width=0.75\linewidth}}  
\put(  110, 165){\text{(c-2)}}
\put(  -5,  -10){\epsfig{file=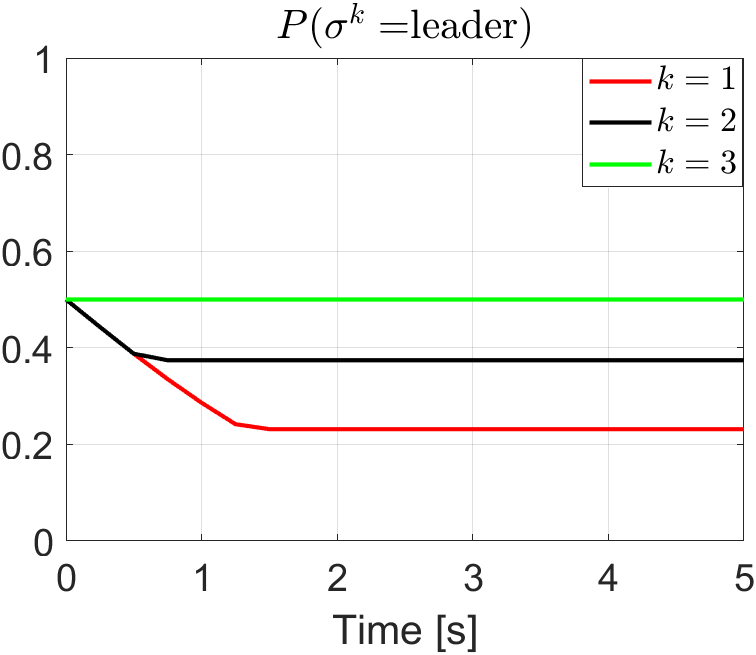,width=0.2\linewidth}}  
\put(  -5, 75){\text{(d-1)}}
\put(  110,  -10){\epsfig{file=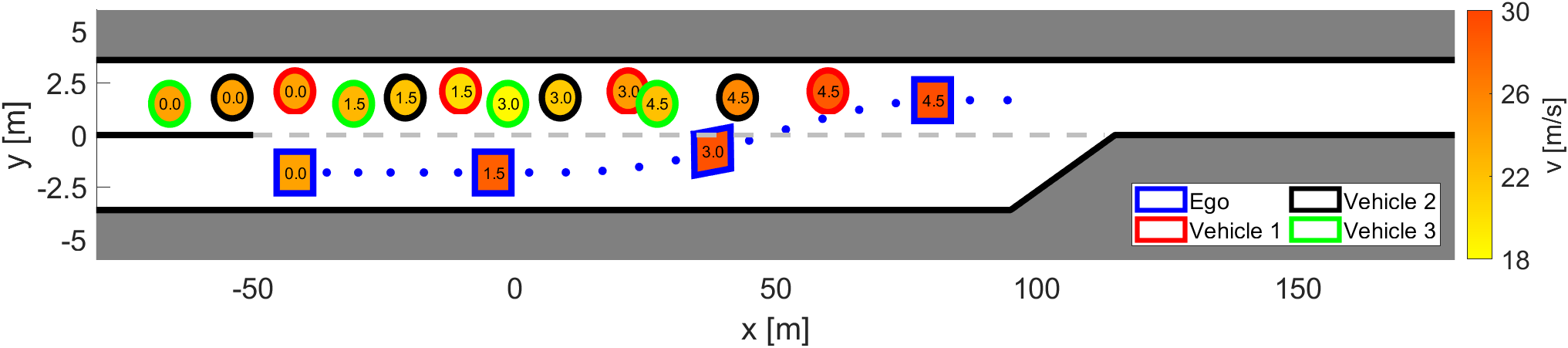,width=0.75\linewidth}}  
\put(  110, 75){\text{(d-2)}}
\end{picture}
\end{center}
      \caption{Results of the proposed LFGC against other drivers following leader/follower policy with different leaders and followers combinations: (a) three leaders; (b) one leader (vehicle 1) and two followers (vehicles 2 and 3); (c) two leaders (vehicles 1 and 2) and one follower (vehicle 3); (d) three followers. The left column (a-1) to (d-1) shows the ego vehicle's belief on other vehicle's being a leader in the game, $\mathbb{P}(\sigma^k = \text{leader}), k = 1, 2, 3$. The right column (a-2) to (d-2) shows the time history results of the ego and other vehicles behaviors during this forced merge process. Specifically, in the right column, the boundary line color of each block distinguishes different vehicles, the number in the block represents time in seconds, the color of each block describes the speed of the vehicle at that time instant, and the blue dotted line represents the ego vehicle's trajectory. Note that vehicles 1-3 share the same vertical positions, and some offsets in vertical direction are added to better distinguish them in the figure.}
      \label{fig:results_LF}
      \vspace{-0.15in}
\end{figure*}
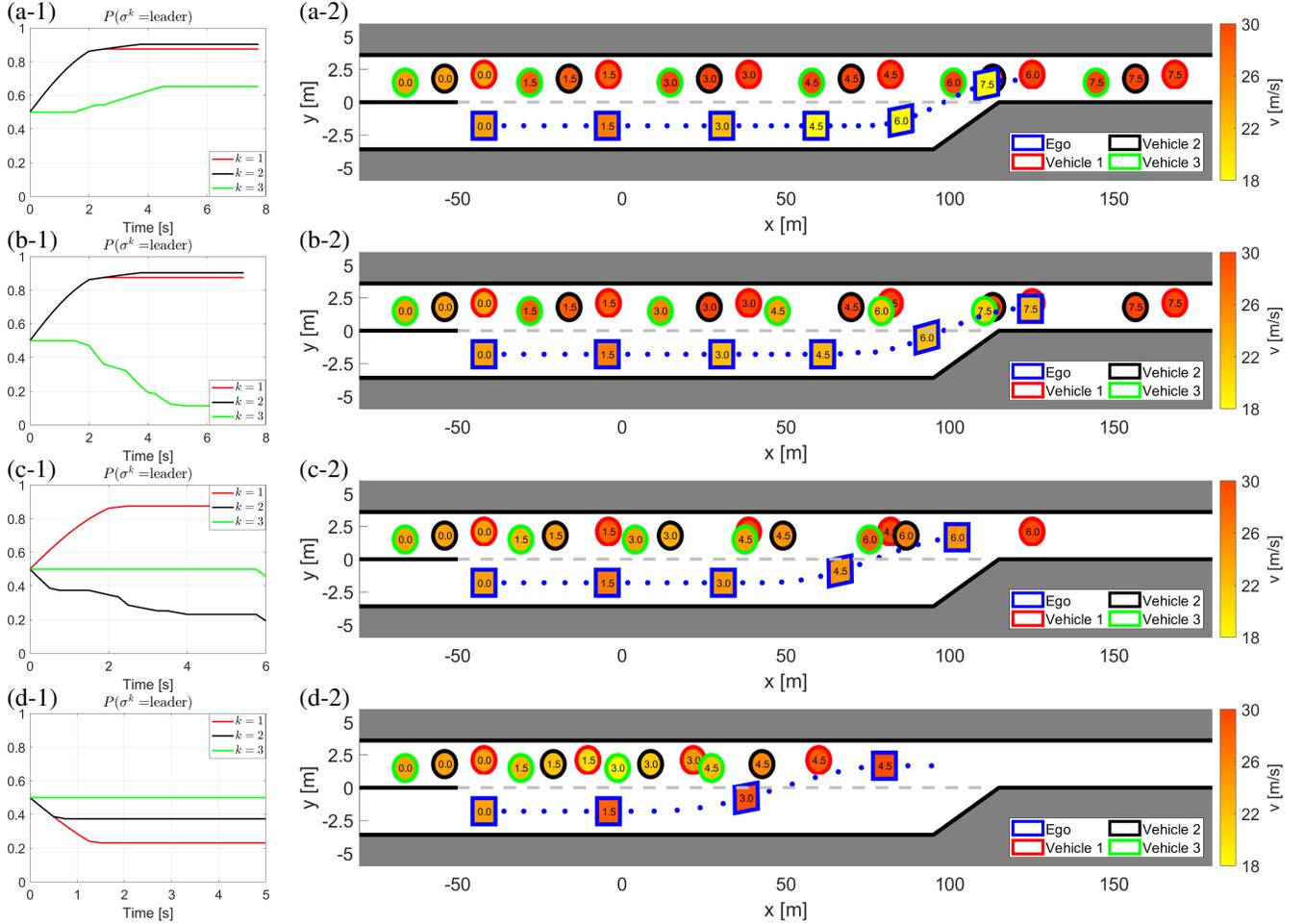

For the LFGC, the planning horizon is selected as $N = 4$ and the chance constraint parameter is chosen as $\varepsilon = 0.1$. Note that a larger $N$ may result in better long-term performance but also lead to longer computational time, while a smaller $N$ may emphasize on immediate benefits and hence fail to merge in many scenarios. For highway forced merge considered in this paper, $N$, in general, needs to be chosen such that it is longer than the duration of the lane change (i.e., $N\Delta T \geq T_{\text{lc}}$). The initial beliefs are set to $\forall k\in \{1,2,3\}, \; \mathbb{P}(\sigma^k = \text{leader}) = \mathbb{P}(\sigma^k = \text{follower}) = 0.5$. Fig.~\ref{fig:results_LF} shows the results when the ego vehicle is interacting with different combinations of leaders and followers. In Fig.~\ref{fig:results_LF}, the left column ((a-1) to (d-1)) shows the ego vehicle belief about each of the other vehicles being a leader in the game, $\mathbb{P}(\sigma^k = \text{leader}), k = 1, 2, 3$. The right column shows the time history results of the ego vehicle and other vehicles behaviors during this forced merge process.

Fig.~\ref{fig:results_LF}(a) shows the results when the ego vehicle is interacting with three leaders. The ego vehicle is able to capture interacting vehicle's intentions that all vehicles are more likely to be leaders in the game as shown in Fig.~\ref{fig:results_LF}(a-1). After obtaining this information, the ego vehicle decides to slow down after $t = 1$ [s] and waits to merge after all interacting vehicles pass. When the ego vehicle is interacting with one leader (vehicle 1) and two followers (vehicle 2 and 3), the ego vehicle recognizes interacting vehicle's intentions correctly as shown in Fig.~\ref{fig:results_LF}(b-1). Then the ego vehicle starts to slow down after $t = 1$ [s], and successfully merges between vehicles 1 and 2, which is shown in Fig.~\ref{fig:results_LF}(b-2). Shown in Fig.~\ref{fig:results_LF}(c) are the results of the ego vehicle interacting with two leaders (vehicles 1 and 2) and one follower (vehicle 3). In this case, the ego vehicle observes that vehicles 1 and 2 speed up and do not yield to it, so the ego vehicle decides to slow down and merge between vehicles 2 and 3. We also perform the test when the ego vehicle interacts with three followers, and the results are shown in Fig.~\ref{fig:results_LF}(d), where the ego vehicle observes all vehicles yielding intentions, speeds up and merges in front of all interacting vehicles. The average computational time for solving \eqref{eq:opt_expect_reward} at each time step is 0.182 [s].

For all cases shown in Fig.~\ref{fig:results_LF}, the initialized beliefs are the same, which means the ego vehicle does not know ahead of time whether the interacting vehicle is a leader or a follower. As a result, the ego vehicle relies on its observations to estimate interacting vehicles leader/follower role. The proposed LFGC can capture interacting vehicles' intentions and making decisions accordingly when all interacting vehicles are controlled by the leader/follower in leader-follower game.

\subsection{Interacting vehicles driven by intelligent driver model} \label{sec:res_idm}

The validation results shown in Section~\ref{sec:res_lf} assumes that other drivers make decisions based on the leader-follower game. The LFGC assumes other drivers are playing leader-follower game with the ego vehicle, estimates their roles in the game, and makes decision accordingly. This means that the environment in Section~\ref{sec:res_lf} behaves just as the LFGC expects. However, the actual behavior of other drivers might be different from leader-follower game's policy. As a result, we want to further investigate how the framework responds to other types of driver models. 

In this section, we employ the intelligent driver model (IDM) to control other vehicles and interact with the ego vehicle. The ego vehicle is still controlled by the LFGC and tries to estimate interacting vehicles' intentions by estimating their corresponding leader or follower roles. IDM is a continuous-time car-following model and is described by \eqref{eq:IDM_1} to \eqref{eq:IDM_3} \cite{treiber2000congested}.
\begin{align}
    &\dot{x} = v, \label{eq:IDM_1}\\
    &\dot{v} = a_m \bigg(1 - \Big(\frac{v}{v_0}\Big)^\delta - \Big(\frac{\phi^*(v,\Delta v)}{\phi}\Big)^2 \bigg), 
\end{align}
where $x$ is the longitudinal position; $v$ is the longitudinal velocity; $v_0$ is the desired velocity of the vehicle; $\phi = x - x_t - l_t$ is the following distance with $x_t$ being the position of the target vehicle and $l_t$ being the length of the target vehicle; $\Delta v = v - v_t$ is the velocity difference of the vehicle and the target vehicle; $\phi^*(v,\Delta v)$ is obtained according to, 
\begin{equation} \label{eq:IDM_3}
    \phi^*(v, \Delta v) = \phi_0 + v T + \frac{v \Delta v}{2\sqrt{a_m b}},
\end{equation}
where $a_m, \phi_0, T, b$ are parameters of the IDM model. The physical interpretation of these parameters are the maximum acceleration $a_m$, the minimum car following distance $\phi_0$, the desired time headway $T$, and the comfortable deceleration $b$.

\begin{figure}[h]
\begin{center}
\begin{picture}(240.0, 50.0)
\put(  -5,  -10){\epsfig{file=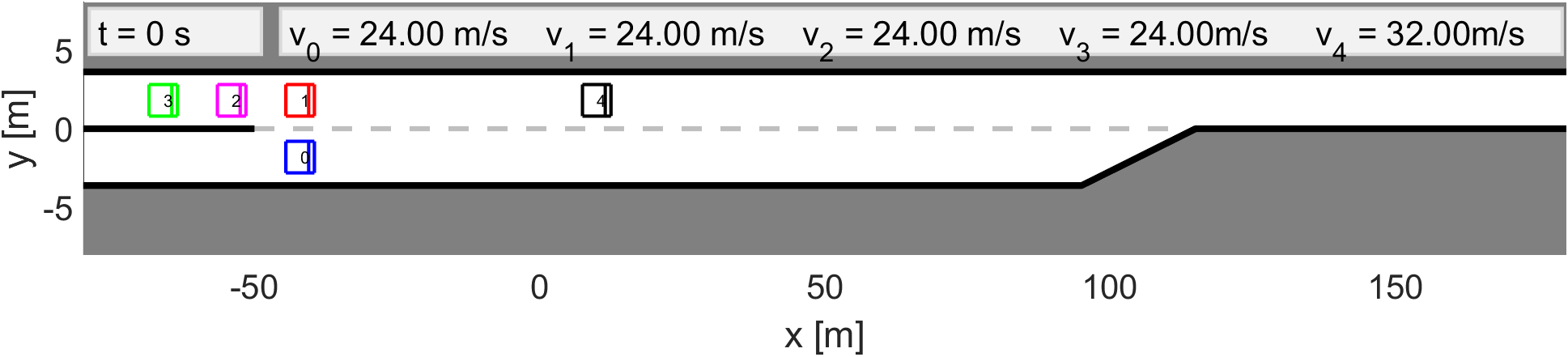,width=1.00\linewidth}}  
\end{picture}
\end{center}
      \caption{Illustration of highway forced merge scenarios for validations of the LFGC when highway vehicles are controlled by IDM.}
      \label{fig:scenario_IDM}
\end{figure}

For the validation tests, we consider the scenario shown in Fig.~\ref{fig:scenario_IDM}. In Fig.~\ref{fig:scenario_IDM}, there is another vehicle ahead of all vehicles (black vehicle 4), and it is driving at a constant speed. The ego vehicle is still the same as in Section~\ref{sec:res_lf} and is controlled by the LFGC, which means that from the ego vehicle perspective, it is playing leader-follower game with all interacting vehicles. For these three interacting vehicles (vehicle 1 to 3), they are controlled by IDM to follow either front vehicle (vehicle 4) or the ego vehicle with certain time headway $T$. The IDM model parameters are listed in Table~\ref{tab:IDM_params}. Note that the ego vehicle regards vehicle 4 as the environmental vehicle and assumes it is driving at constant speed.

Different desired time headway in IDM may reflect conservativeness of the drivers. If the interacting vehicle intends to yield to the ego vehicle, we model it to use IDM  to follow the ego vehicle with certain time headway. This means each interacting vehicle has an option to follow either the front vehicle or the ego vehicle.

\begin{table}[h!]
\centering
\begin{tabular}{ |c||c| } 
 \hline
 Parameters & Values \\ 
 \hline
 \hline
 Desired velocity $v_0$ & $32$ m/s  \\ 
 Minimum spacing $\phi_0$ & $2$ m \\
 Maximum acceleration $a_m$ & $4$ m/s$^2$ \\
 Comfortable deceleration $b$ & $3$ m/s$^2$ \\
 Acceleration exponent $\delta$ & $4$ \\
 Desired time headway $T$ & $0.5$ to $2.5$ s \\
 \hline
\end{tabular}
\caption{Intelligent driver model parameters.}
\label{tab:IDM_params}
\vspace{-0.2in}
\end{table}

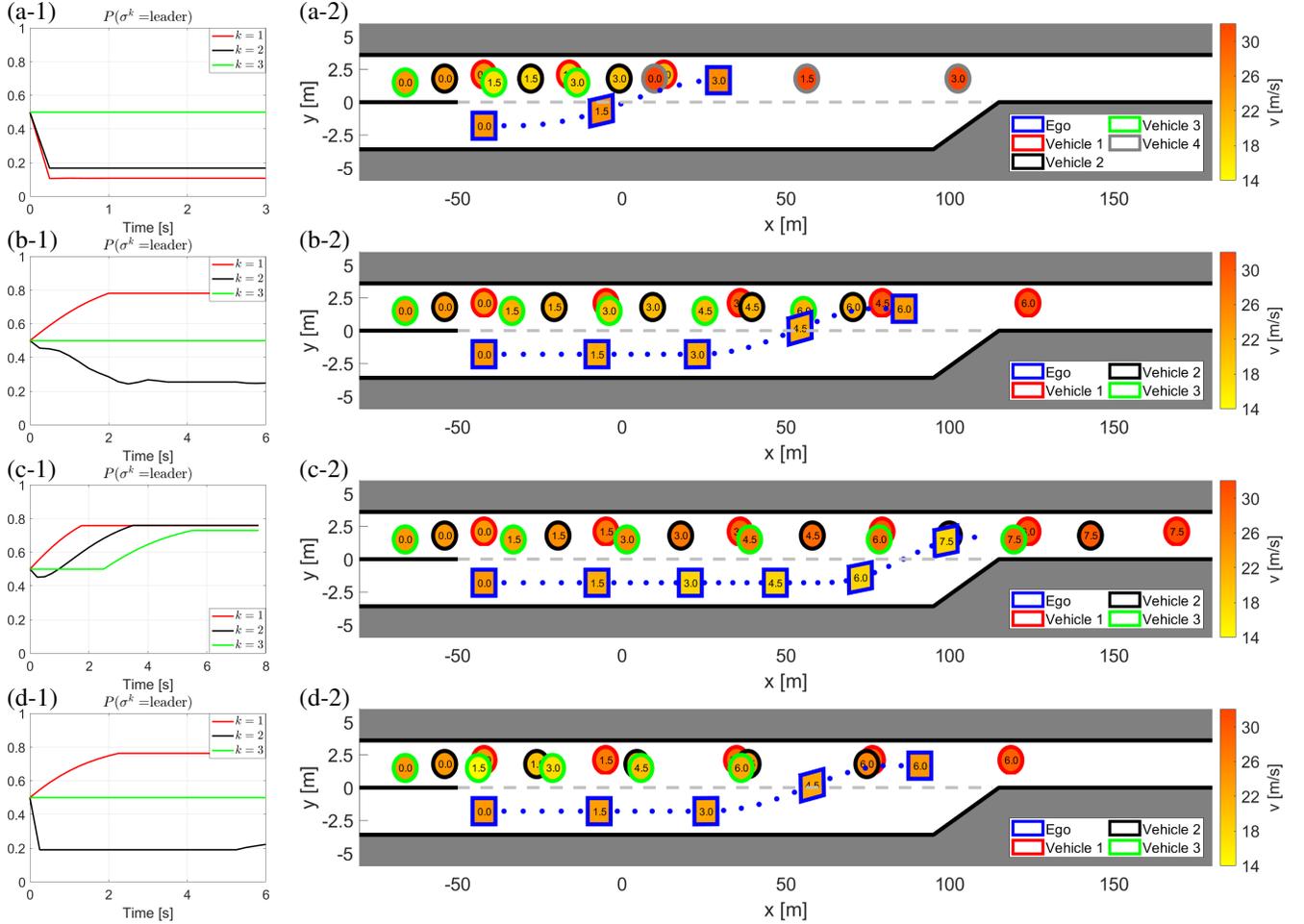
\begin{figure*}[t]
\begin{center}
\begin{picture}(480.0, 360.0)
\put(  -5,  260){\epsfig{file=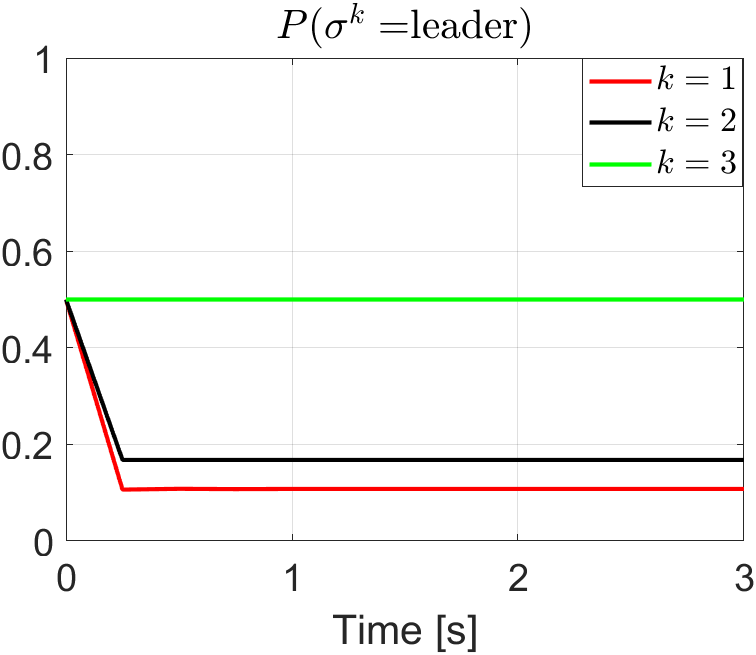,width=0.2\linewidth}}  
\put(  -5, 345){\text{(a-1)}}
\put(  110,  260){\epsfig{file=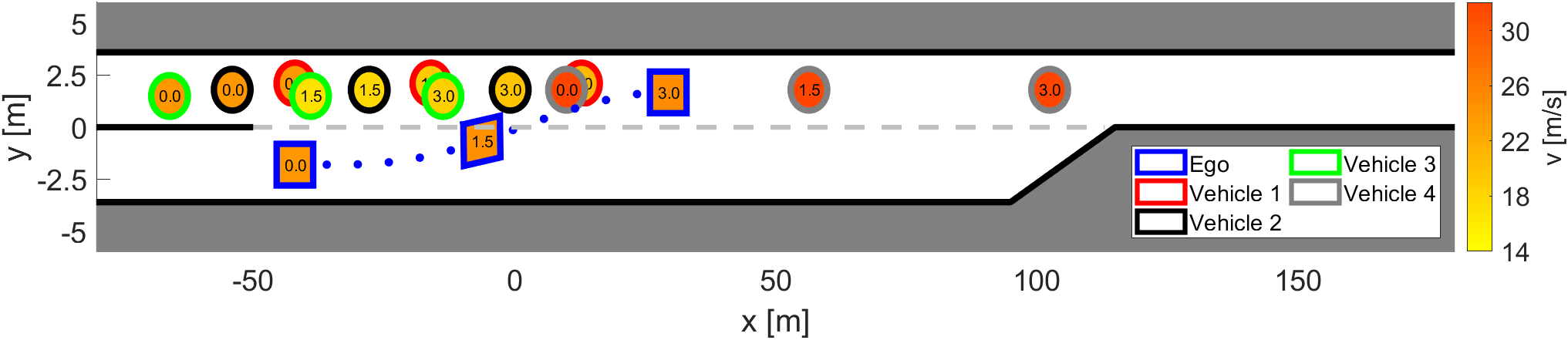,width=0.75\linewidth}}  
\put(  110, 345){\text{(a-2)}}
\put(  -5,  170){\epsfig{file=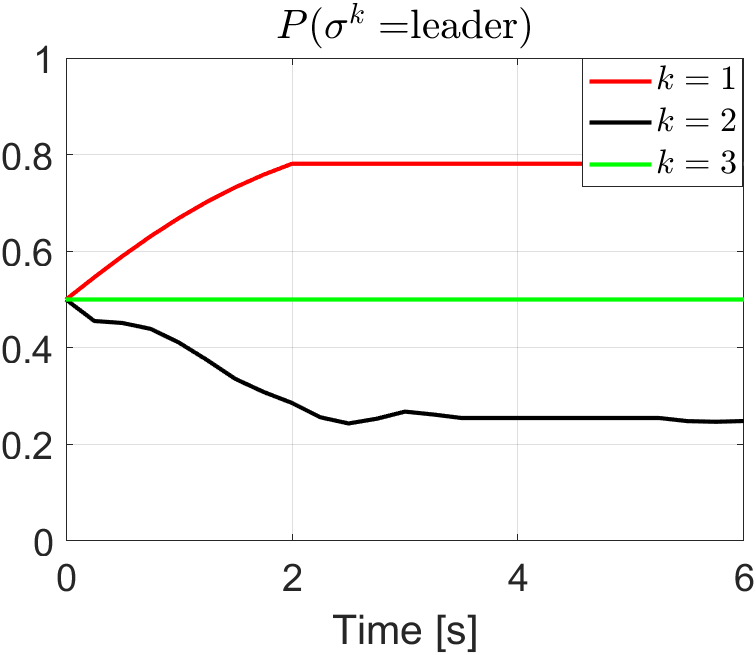,width=0.2\linewidth}}  
\put(  -5, 255){\text{(b-1)}}
\put(  110,  170){\epsfig{file=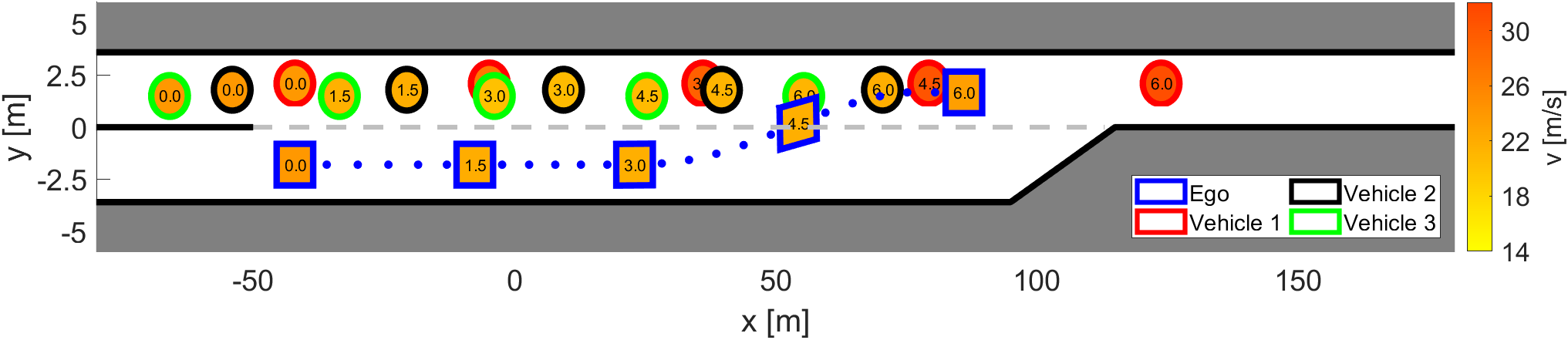,width=0.75\linewidth}}  
\put(  110, 255){\text{(b-2)}}
\put(  -5,  80){\epsfig{file=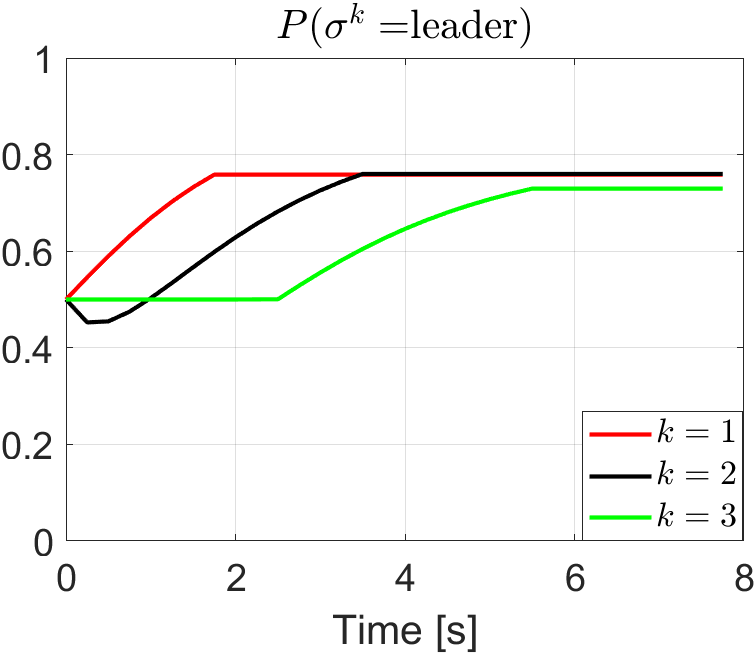,width=0.2\linewidth}}  
\put(  -5, 165){\text{(c-1)}}
\put(  110,  80){\epsfig{file=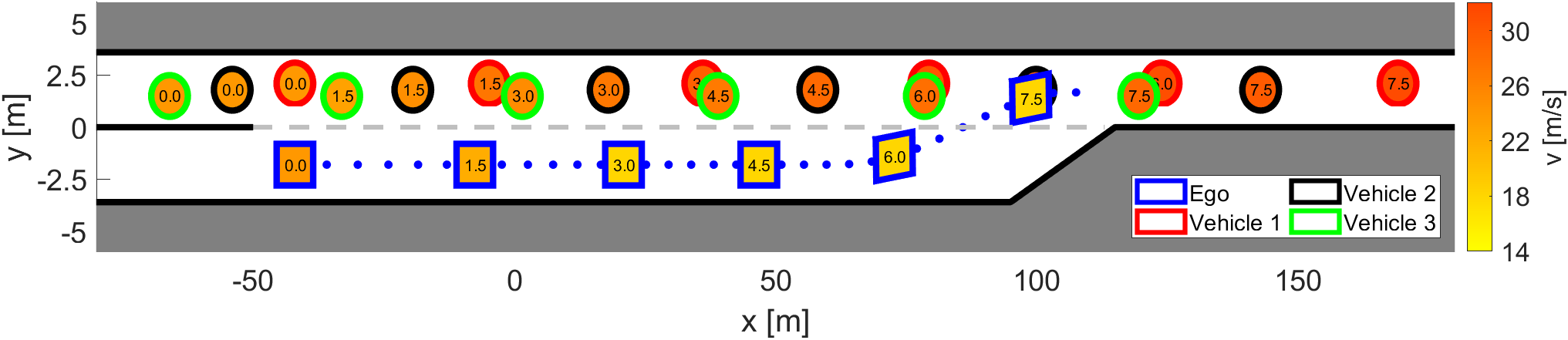,width=0.75\linewidth}}  
\put(  110, 165){\text{(c-2)}}
\put(  -5,  -10){\epsfig{file=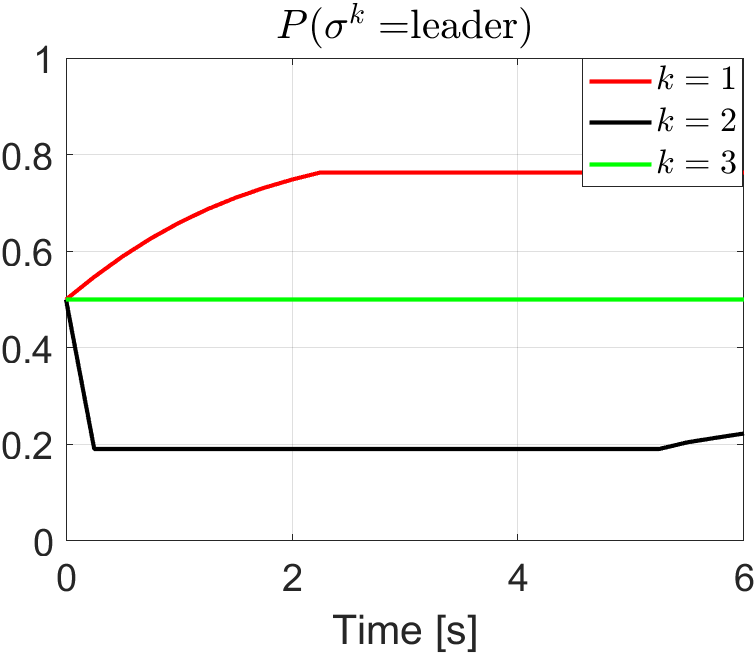,width=0.2\linewidth}}  
\put(  -5, 75){\text{(d-1)}}
\put(  110,  -10){\epsfig{file=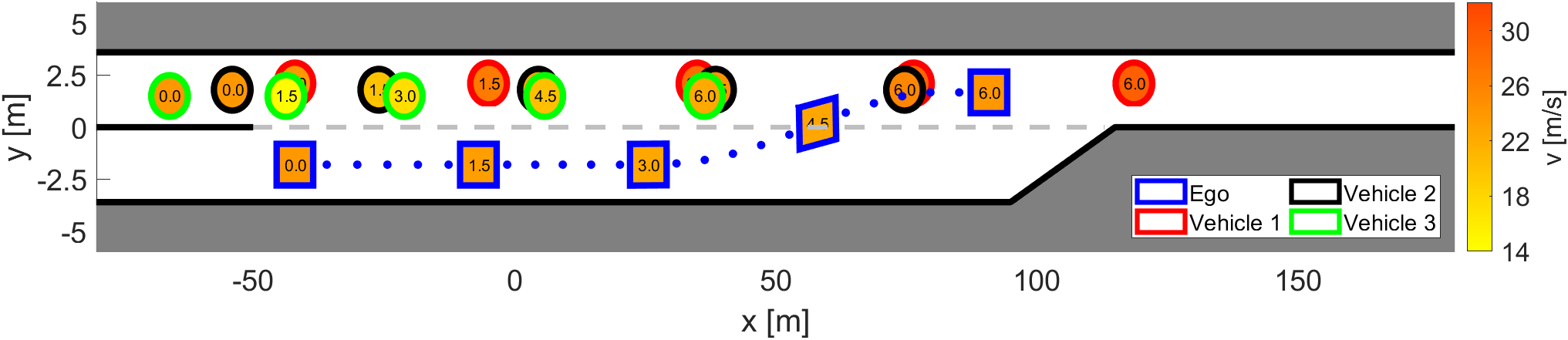,width=0.75\linewidth}}  
\put(  110, 75){\text{(d-2)}}
\end{picture}
\end{center}
      \caption{Results of the LFGC against other vehicles controlled by IDM with different target vehicles and desired time headway: (a) vehicle 1 yields (follows the ego vehicle) with time headway $T=1$ [s] and vehicles 2 and 3 follow the front vehicle with $T=0.5$ [s]; (b) vehicle 2 yields (follows the ego vehicle) with $T=0.5$ [s] and vehicles 1 and 3 follow the front vehicle with $T=0.5$ [s]; (c) all vehicles follow the front vehicle with $T = 0.5$ [s]; (d) all vehicles follow front vehicle with $T = 1.5$ [s]. The left column (a-1) to (d-1) shows the ego vehicle's belief on other vehicle's being a leader in the game, $\mathbb{P}(\sigma^k = \text{leader}), k = 1, 2, 3$. The right column (a-2) to (d-2) shows the time history results of the ego and other vehicles behaviors during this forced merge process. Specifically, in the right column, the boundary line color of each block distinguishes different vehicles, the number in the block represents time in seconds, the color of each block describes the speed of the vehicle at that time instant, and the blue dotted line represents the ego vehicle's trajectory. Note that vehicles 1-4 share the same vertical positions, and some offsets in vertical direction are added to better distinguish them in the figure. Note also that since vehicle 4 (grey boundary) drives at constant speed, the time history results of vehicle 4 is only shown in (a) but omitted in (b) to (d) for clarity.}
      \label{fig:results_IDM}
      \vspace{-0.15in}
\end{figure*}

For the LFGC, the setting is the same as in Section~\ref{sec:res_lf}. The planning horizon is selected as $N = 4$ and the chance constraint parameter is chosen as $\varepsilon = 0.1$. The initial beliefs are set to $\forall k\in \{1,2,3\}, \; \mathbb{P}(\sigma^k = \text{leader}) = \mathbb{P}(\sigma^k = \text{follower}) = 0.5$. Fig.~\ref{fig:results_IDM} shows the results when the ego vehicle is interacting with other vehicles controlled by IDM with different target vehicles and different desired time headway.

In Fig.~\ref{fig:results_IDM}(a), the first interacting vehicle (vehicle 1) intends to yield to the ego vehicle, so it chooses to follow the ego vehicle with 1 [s] time headway, while the last two interacting vehicles are following the front vehicles with 0.5 [s] headway. From Fig.~\ref{fig:results_IDM}(a-1), the ego vehicle thinks that vehicle 1 has a high probability being a follower in the game and chooses to merge in front of vehicle 1 as depicted in Fig.~\ref{fig:results_IDM}(a-2). Fig.~\ref{fig:results_IDM}(b) shows another case where the first interacting vehicle (vehicle 1) follows the front vehicle with 0.5 [s] headway, and the second interacting vehicle intends to yield to the ego vehicle and follows the ego vehicle with 0.5 [s] headway. Then in this case, from the ego vehicle perspective, vehicle 1 has higher probability being a leader in the game, while vehicle 2 has a higher probability being a follower in the game, and hence the ego vehicle successfully merges in front of vehicle 2 in this case. Two other non-yield cases are shown in Fig.~\ref{fig:results_IDM}(c) and (d). Fig.~\ref{fig:results_IDM}(c) shows the results of all interacting vehicles following the front vehicle with 0.5 [s] headway. From the ego vehicle perspective, all interacting vehicles are more likely to be leaders in the game, so the ego vehicle successfully merges after all vehicles pass. In Fig.~\ref{fig:results_IDM}(d), all interacting vehicles follow the front vehicle with 1.5 [s] headway. In this case, the ego vehicle finds that vehicle 2’s behavior is conservative and thinks vehicle 2 has a higher probability being a follower in the game. Hence, the ego vehicle successfully merges between vehicles 1 and 2. The average computational time for solving \eqref{eq:opt_expect_reward} at each time step is 0.198 [s].

For all cases shown in Fig.~\ref{fig:results_IDM}, the ego vehicle starts with the same initial belief. This means that the ego vehicle does not know other drivers conservativeness (represented by desired time headway) and intentions (represented by target vehicles) a priori. The ego vehicle relies on the LFGC to estimate their intentions, make decisions accordingly and is able to merge successfully.

\subsection{Interacting vehicles following traffic data} \label{sec:res_us101}

We have already tested the LFGC with other vehicles driven by leader/follower in the leader-follower game and by IDM models. We want to further test the controller's performance against real traffic data. Specifically, we use the US highway 101 traffic dataset from the Next Generation Simulation (NGSIM) website \cite{us101data}, which is collected by the United States Federal Highway Administration and is considered as one of the largest publicly available sources of naturalistic driving data. The US highway 101 dataset has been extensively studied in the literature \cite{lu2007freeway, altche2017lstm, deo2018multi}.

\begin{figure}[h]
\begin{center}
\begin{picture}(240.0, 165.0)
\put(  25,  -10){\epsfig{file=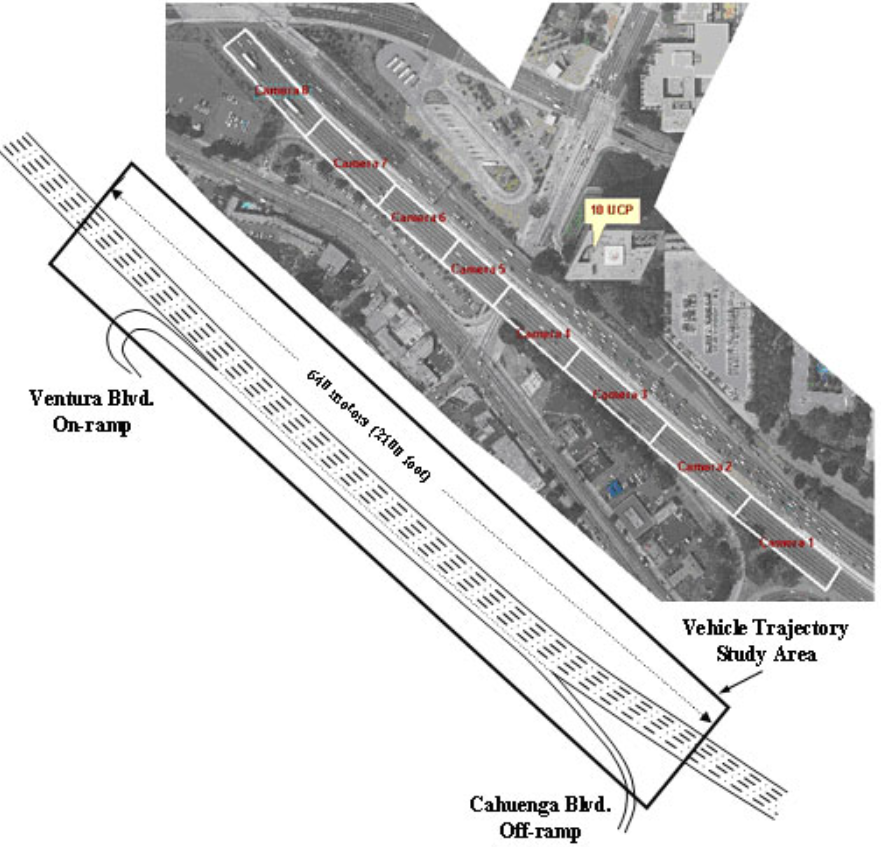,width=0.7\linewidth}}  
\end{picture}
\end{center}
      \caption{Top view of the highway that is used for collecting the US101 traffic data \cite{us101data}. The section of interest includes five main lanes of the highway, one on-ramp to the highway, one off-ramp exiting the highway, and also one auxiliary lane that is used to merge into the highway and exit the highway.}
      \label{fig:US101_topview}
\end{figure}

More specifically, we consider a portion of the US101 traffic dataset that contains 30 minutes of vehicles trajectories on the US101 highway. The time period ranges from 7:50 to 8:20 am, which represents the buildup of congestion around the morning peak hours. The dataset contains position and velocity trajectories as well as vehicle dimensions for around 6000 vehicles, and the information is recorded every 0.1 [s]. The top view of the portion of the US101 highway that is used for collecting the data is shown in Fig.~\ref{fig:US101_topview}. The studied section consists of five main lanes of the highway, one on-ramp to the highway, one off-ramp exiting the highway, and also one auxiliary lane that is used to merge into the highway and exit the highway. 

\begin{figure}[h]
\begin{center}
\begin{picture}(240.0, 115.0)
\put(  25,  -10){\epsfig{file=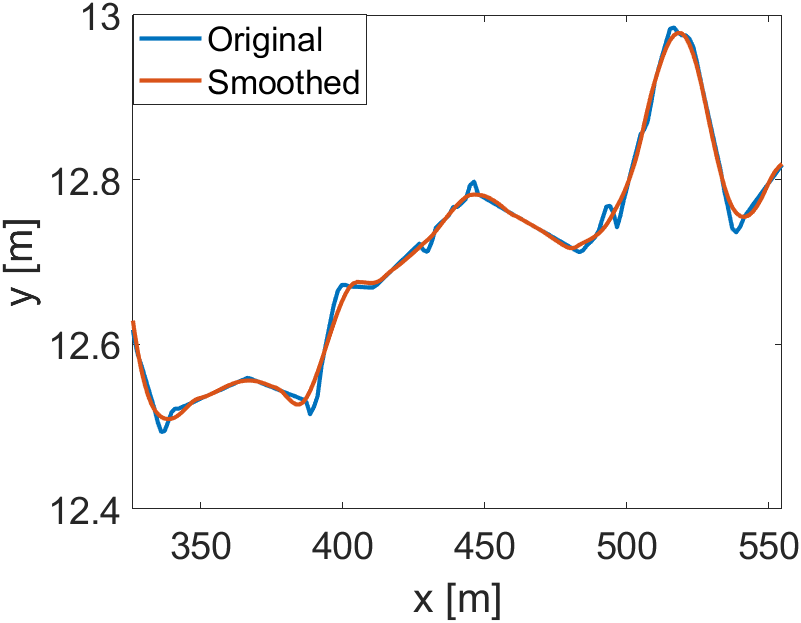,width=0.7\linewidth}}  
\end{picture}
\end{center}
      \caption{Smooth vehicle trajectories from the US101 traffic dataset using the Savitzky-Golay filter.}
      \label{fig:smooth_traj}
\end{figure}

As discussed in \cite{montanino2013making}, the US101 dataset contains a significant amount of noise due to video analysis and numerical differentiation. To overcome this drawback, the Savitzky-Golay filter \cite{savitzky1964smoothing} is utilized to smooth vehicles' positions and update their corresponding velocities. The Savitzky-Golay filter performs well for signal differentiation and smoothing the US101 dataset with window length 21 \cite{altche2017lstm}. One original vehicle trajectory and the corresponding smoothed vehicle trajectory are shown in Fig.~\ref{fig:smooth_traj}. 

\begin{figure}[h]
\begin{center}
\begin{picture}(240.0, 55.0)
\put(  -5,  -10){\epsfig{file=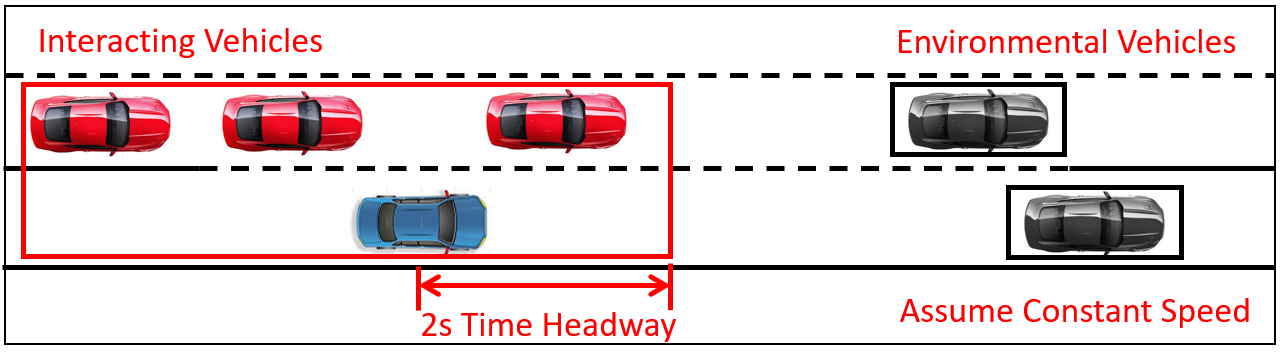,width=1.0\linewidth}}  
\end{picture}
\end{center}
      \caption{Selection of interacting vehicles: The ego vehicle (blue vehicle) considers vehicles inside the selection box (red box) as the interacting vehicles. The front end of the selection box is 2 [s] time headway in front of the ego vehicle. The first vehicle in the target lane within the selection box is regarded as the first interacting vehicle, and the following vehicles are regarded as second and third interacting vehicles. For all other vehicles on the highway, they are treated as environmental vehicles and are assumed to maintain a constant speed.}
      \label{fig:us101_intselection}
\end{figure}

\begin{figure}[H]
\begin{center}
\begin{picture}(240.0, 90.0)
\put(  -5,  -10){\epsfig{file=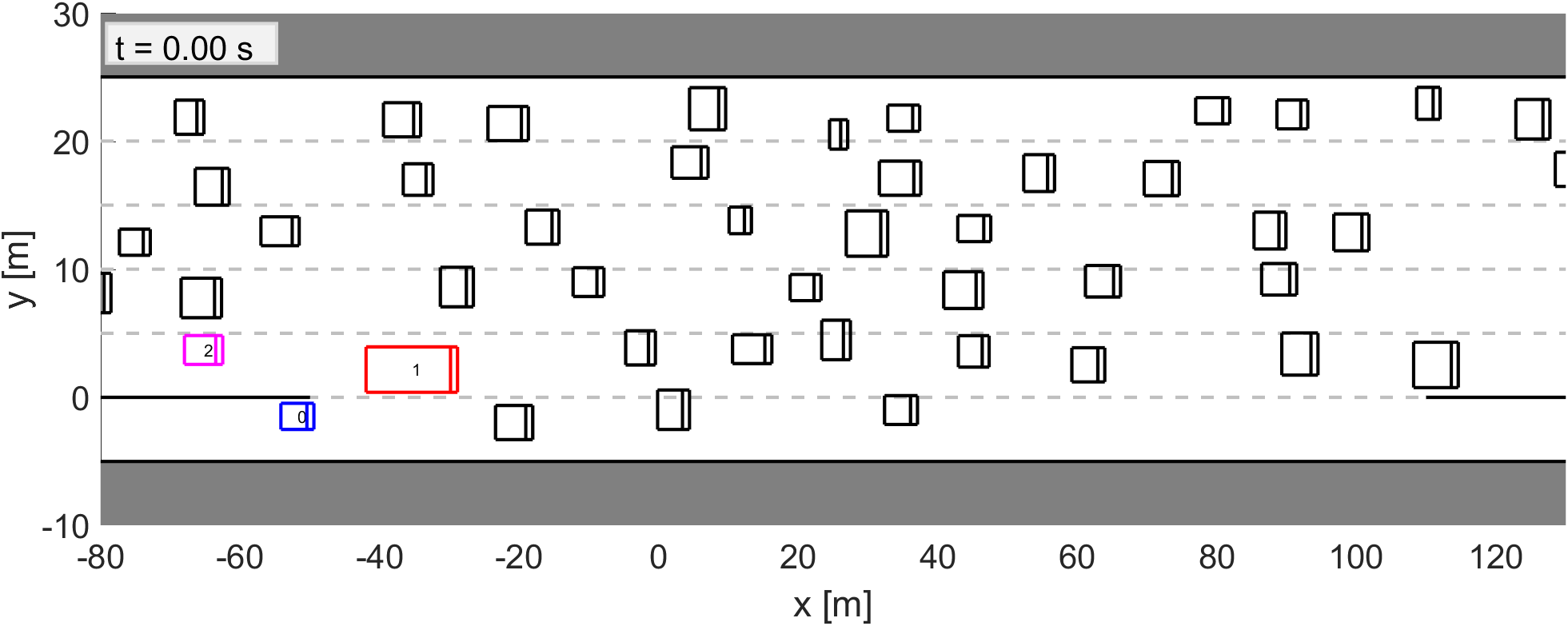,width=1.0\linewidth}}  
\end{picture}
\end{center}
      \caption{One merge scenario identified from the US101 traffic dataset. In this scenario, vehicle 0 (blue vehicle) is the merge vehicle, and we let the LFGC to control vehicle 0. Based on our criteria of selecting interacting vehicles, vehicle 1 (red vehicle) and vehicle 2 (pink vehicle) are selected as the interacting vehicles, and all other vehicles (black vehicle) are regarded as environmental vehicles, which are assumed to drive at constant speed.}
      \label{fig:us101_scene}
\end{figure}

For the validation tests of the LFGC, we focus on the on-ramp and the auxiliary lane to identify all merging vehicles. After identifying the merging vehicles and the corresponding scenario, we identify the interacting vehicles according to Fig.~\ref{fig:us101_intselection}. Specifically, we consider the first vehicle in the target lane that is within 2 [s] time headway in front of the ego vehicle as the first interacting vehicle and regard the consecutive vehicles as the second and third vehicles. For all other vehicles present in the scenario, the ego vehicle will regard them as environmental vehicles and assume they drive at constant speed. One identified merging scenario is shown in Fig.~\ref{fig:us101_scene}.

For each merging scenario, instead of letting the ego vehicle to follow the traffic data, we use the LFGC to control ego vehicle's action and resulting trajectory. For all other vehicles including interacting vehicles and environmental vehicles, they follow their corresponding trajectories present in the US101 traffic dataset. Then, the LFGC needs to estimate interacting vehicles' intentions and control the ego vehicle to merge appropriately. Note that interacting vehicles and environmental vehicles may interact with the merging vehicle in the actual traffic during the data collection. Since the LFGC may take different actions from human's operation, interacting vehicles' or environmental vehicles' behaviors are not responding to the ego vehicle's action. Instead, their behaviors are pre-determined by the traffic dataset.

\begin{table}[h!]
\centering
\begin{tabular}{ |c||c|c|c| } 
 \hline
 Time: 06/15/2015 & 7:50 to 8:05 am & 8:05 to 8:20 am & Total\\ 
 \hline
 \hline
 Number of Merges& 119 & 79 & 198  \\ 
 Success & 116 & 77 & 193 \\
 Fail to Merge & 2 & 2 & 4\\
 Collision & 1 & 0 & 1\\
 Success Rate & 97.5\% & 97.5\% & 97.5\%\\
 \hline
\end{tabular}
\caption{Statistics of validating the LFGC using the US101 traffic dataset. "Success" means the ego vehicle successfully merges to the target lane without any collisions. "Fail to Merge" means that the ego vehicle fails to merge by the end of the auxiliary lane. "Collide" means the ego vehicle collides with other vehicles. }
\label{tab:us101_stats}
\end{table}

Statistics of validating the LFGC based on the US101 traffic dataset are shown in Table~\ref{tab:us101_stats}. There are a total of 198 merge cases present in the dataset that happen from 7:50 to 8:20 am. The average computational time for solving \eqref{eq:opt_expect_reward} at each time step among all merge cases is 0.259 [s]. In 193 merge cases, the LFGC successfully maneuvers the ego vehicle to merge to the target lane. The LFGC fails in 5 cases including 4 "Fail to Merge" cases and 1 "Collision" case. The reason for these failure cases is primarily due to either 1) the LFGC cannot obtain a high belief on interacting vehicles' driving intentions and hence needs to take conservative action to avoid collision, or 2) the traffic is dense such that there is no safe margin for the ego vehicle to merge without intersecting with other vehicles' collision boxes.

\begin{figure}[h]
\begin{center}
\begin{picture}(240.0, 530.0)
\put(  -10,  425){\epsfig{file=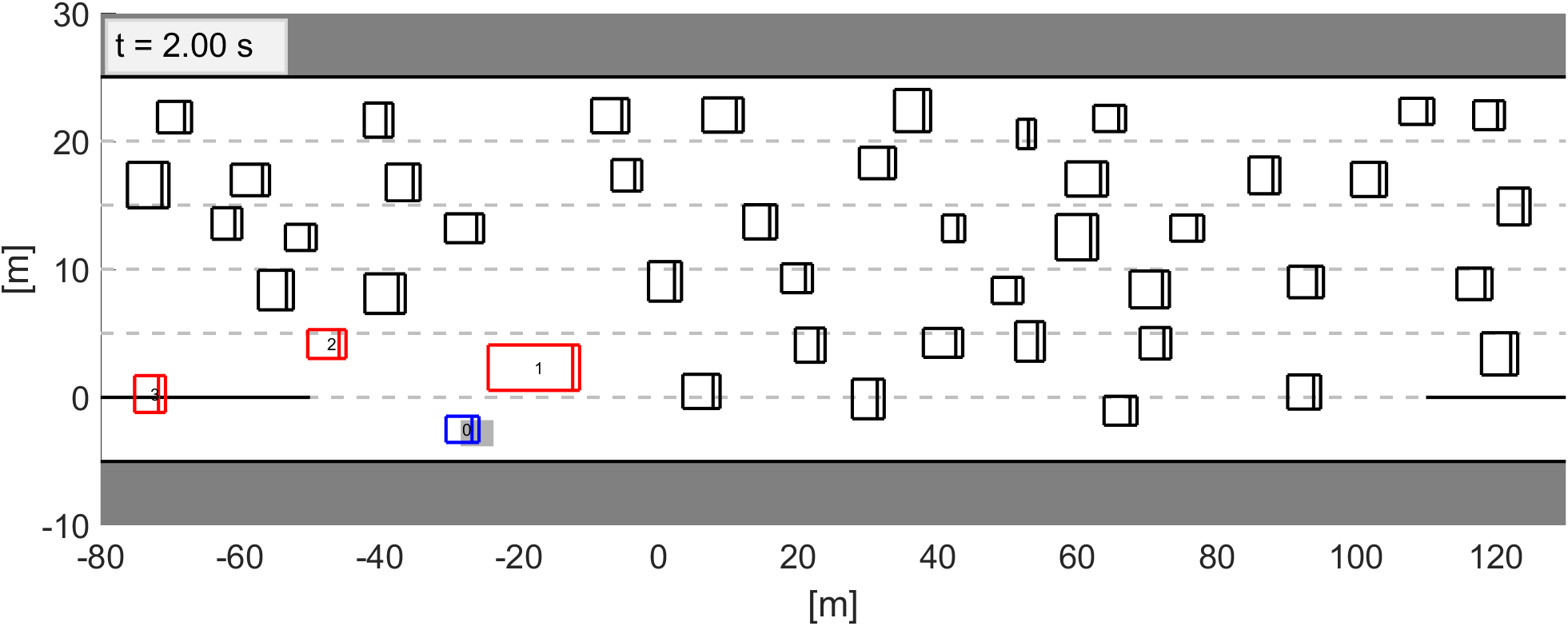,width=1.00\linewidth}}  
\put(  -10,  338){\epsfig{file=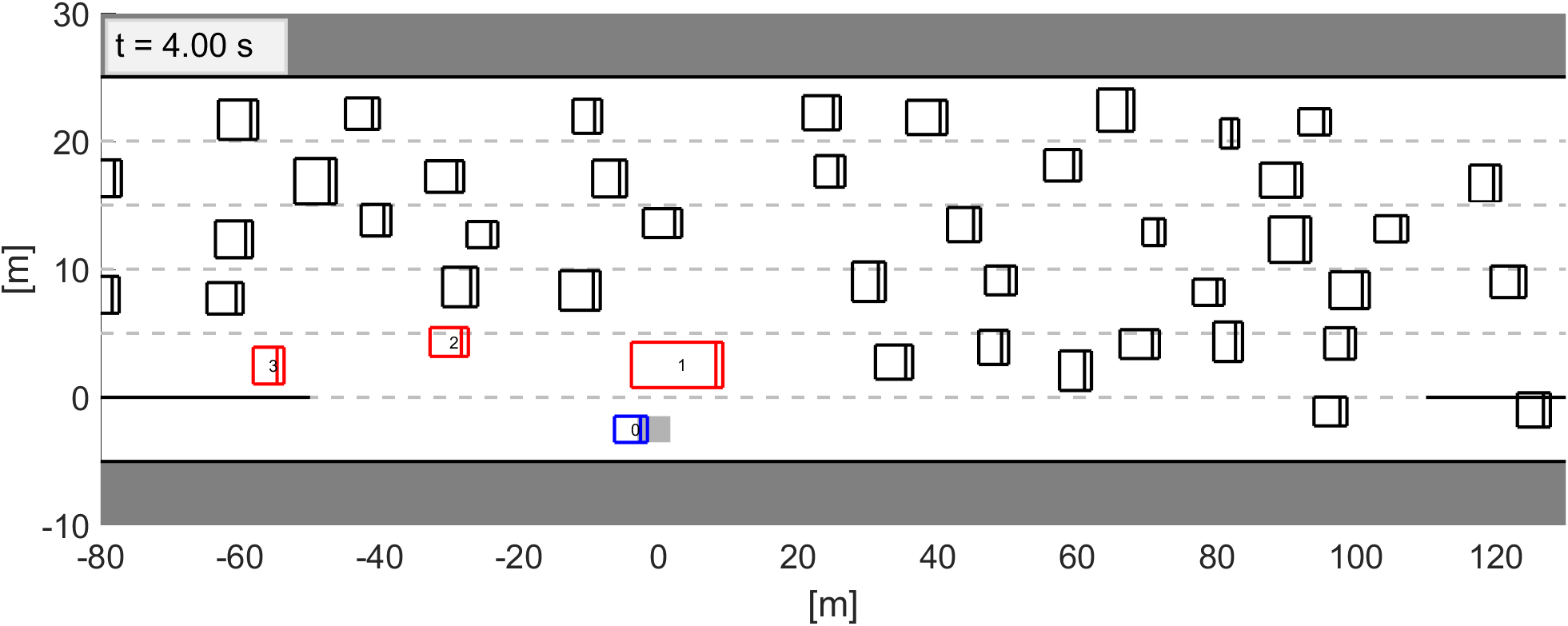,width=1.00\linewidth}}  
\put(  -10,  251){\epsfig{file=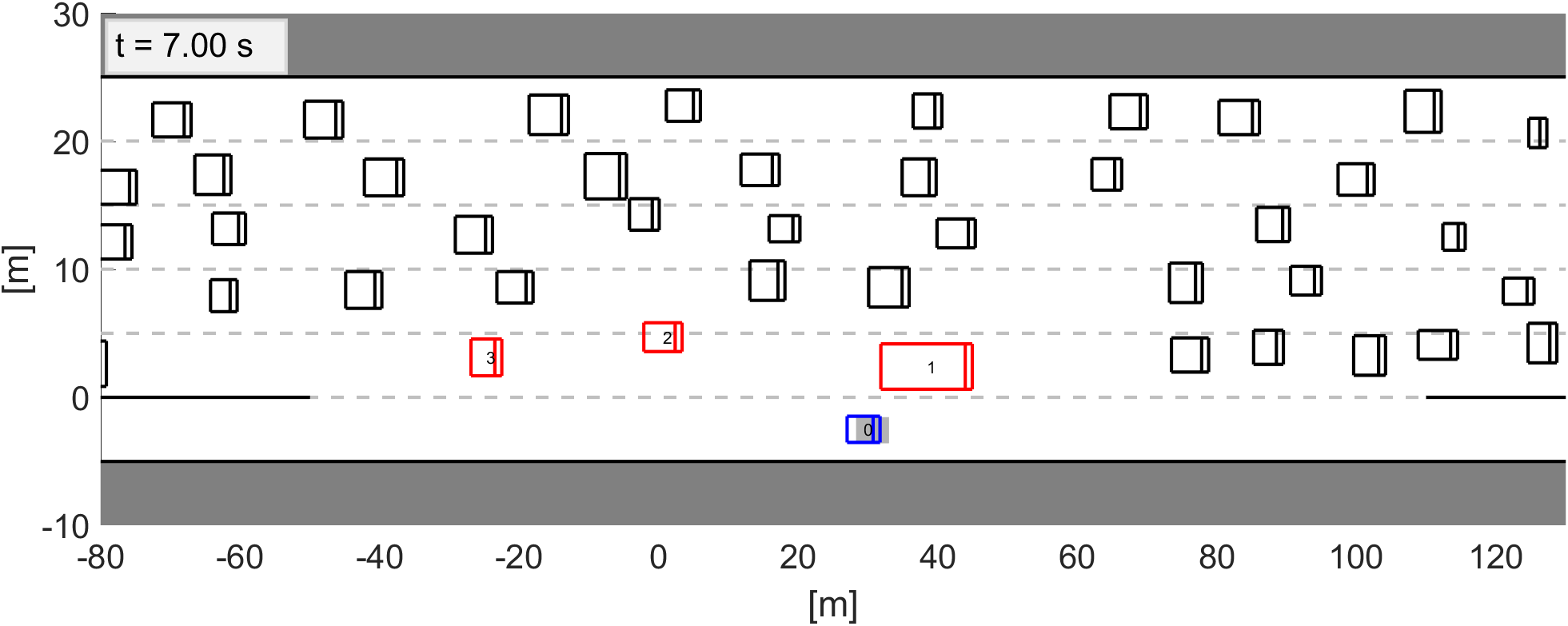,width=1.00\linewidth}}  
\put(  -10,  164){\epsfig{file=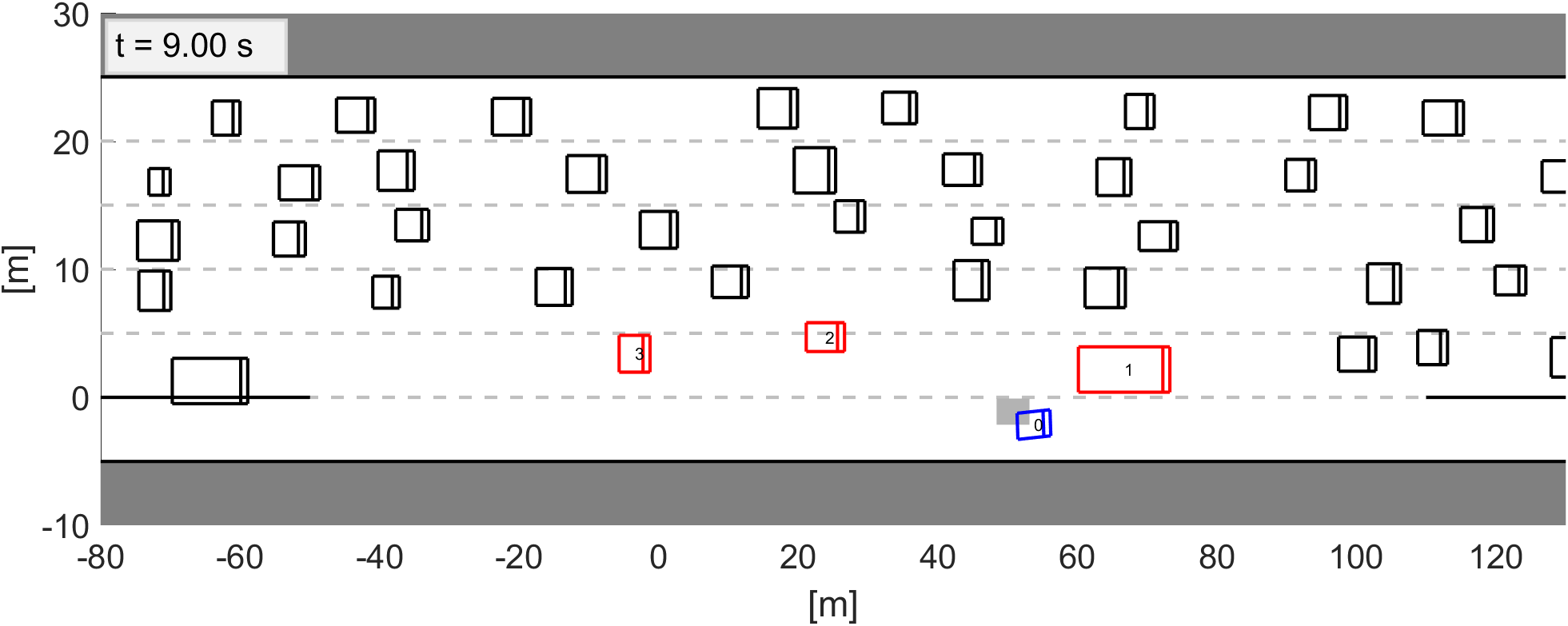,width=1.00\linewidth}}  
\put(  -10,  77){\epsfig{file=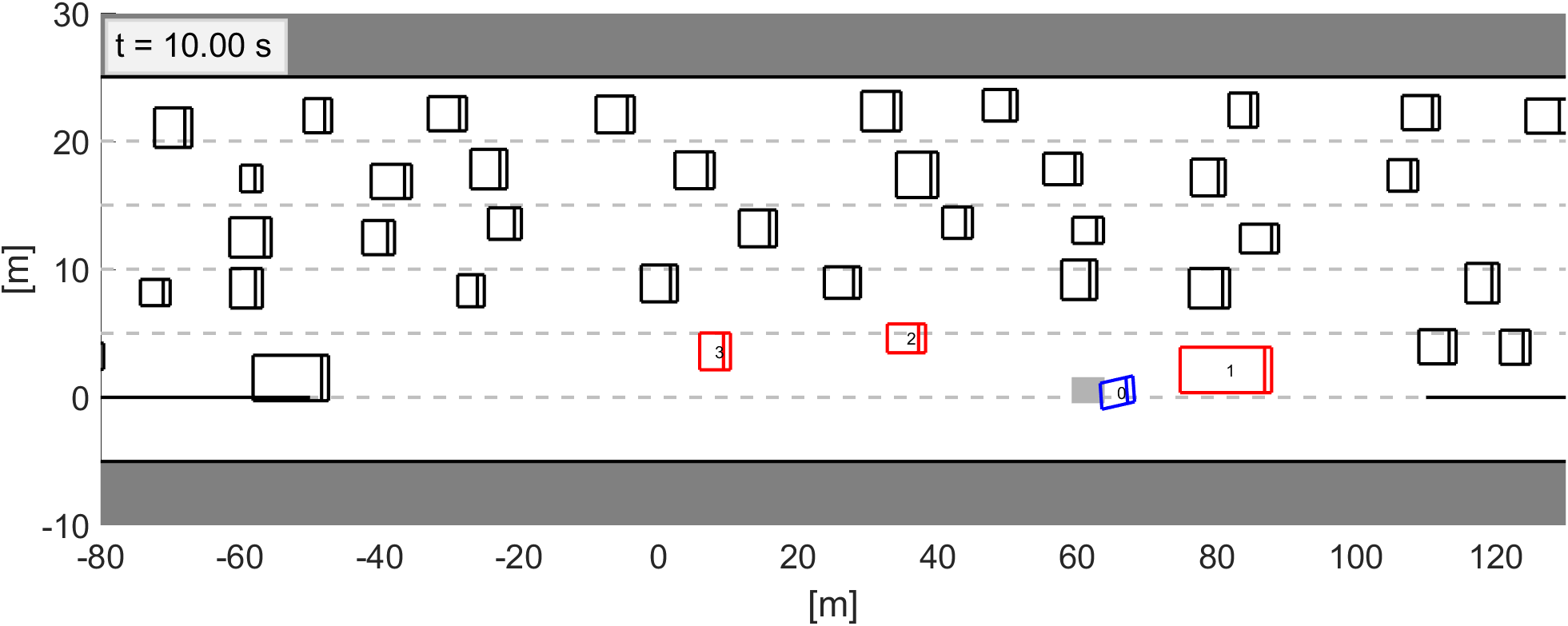,width=1.00\linewidth}}  
\put(  -10,  -10){\epsfig{file=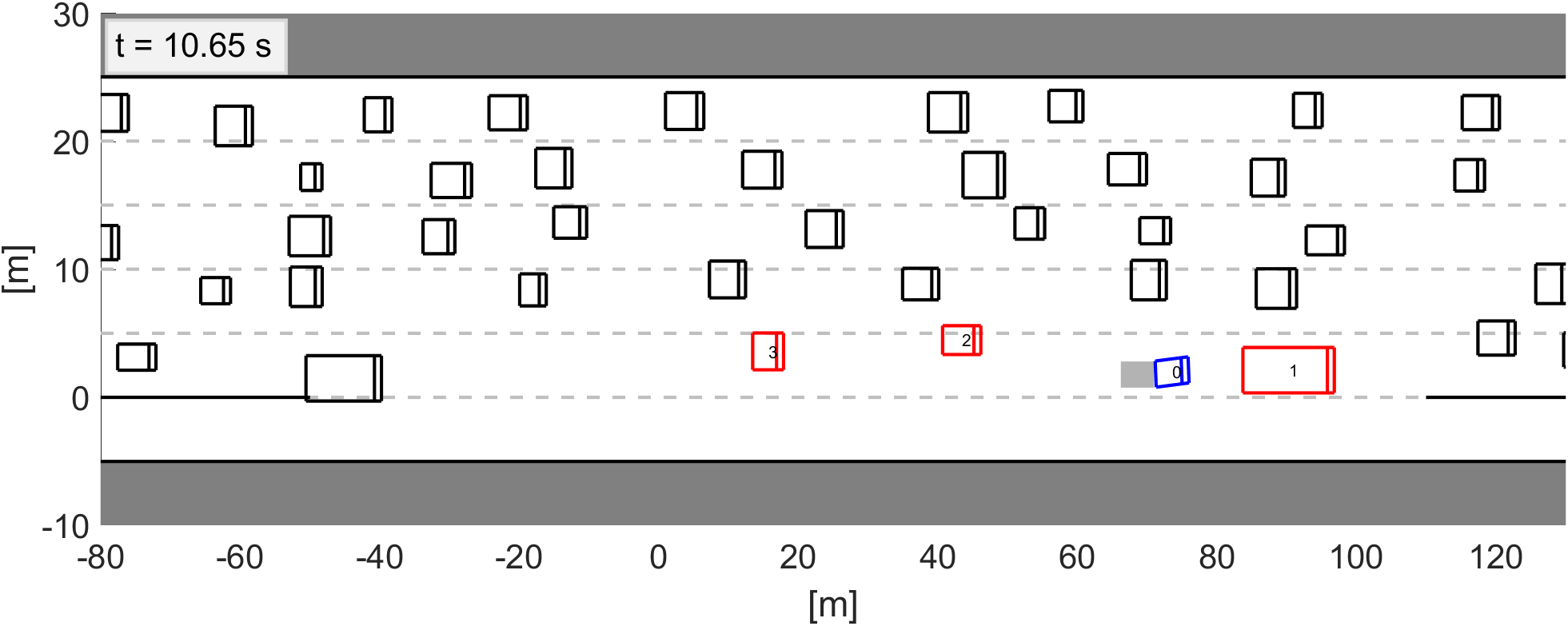,width=1.00\linewidth}}  
\end{picture}
\end{center}
      \caption{An illustration of a successful merge when validating the LFGC against the US Highway 101 dataset. The blue vehicle is the ego vehicle controlled by the LFGC, and the grey box is the position of the ego vehicle present in the dataset.}
      \label{fig:us101_success}
\end{figure}

In Fig.~\ref{fig:us101_success}, we present screenshots for one successful merges. In these figures, the blue vehicle is controlled by the LFGC, and the grey box represent the actual position of the ego vehicle in the dataset. All other vehicles (including red interacting vehicles and black environmental vehicles) are following their corresponding trajectories in the dataset. The ego vehicle controlled by the LFGC makes similar decisions compared to the human driver (grey box): Both the LFGC and the human driver try to speed up and merge in front of the truck (Vehicle 1) at first. However, after recognizing that the truck is more likely to proceed without yielding, the ego vehicle decides to slow down and merges after the truck.

%% file: sections/06_conclusion.tex
\section{Summary} \label{sec:summary}

In this paper, we proposed a Leader-Follower Game Controller (LFGC) for autonomous vehicle planning and control in merge scenarios. The LFGC treats interaction uncertainties due to different driver intentions as latent variables, estimates on-board other driver intentions, and chooses actions to facilitate ego vehicle's merge. In particular, the LFGC is able to perform a receding horizon optimization subject to an explicit probabilistic safety characterization i.e., subject to constraints representing vehicle safety requirements. By considering pairwise interactions of the ego vehicle and interacting vehicles, the LFGC is able to handle interactions with multiple vehicles in a computational tractable way. Finally, multiple simulation-based validations are performed to demonstrate effectiveness of the LFGC, including scenarios that other vehicles are following leaders or followers in the game, the Intelligent Driver Model (IDM), and actual US Highway 101 data.

%% file: sections/07_biography.tex
\begin{IEEEbiography}[{\includegraphics[width=1in,height=1.25in,clip,keepaspectratio]{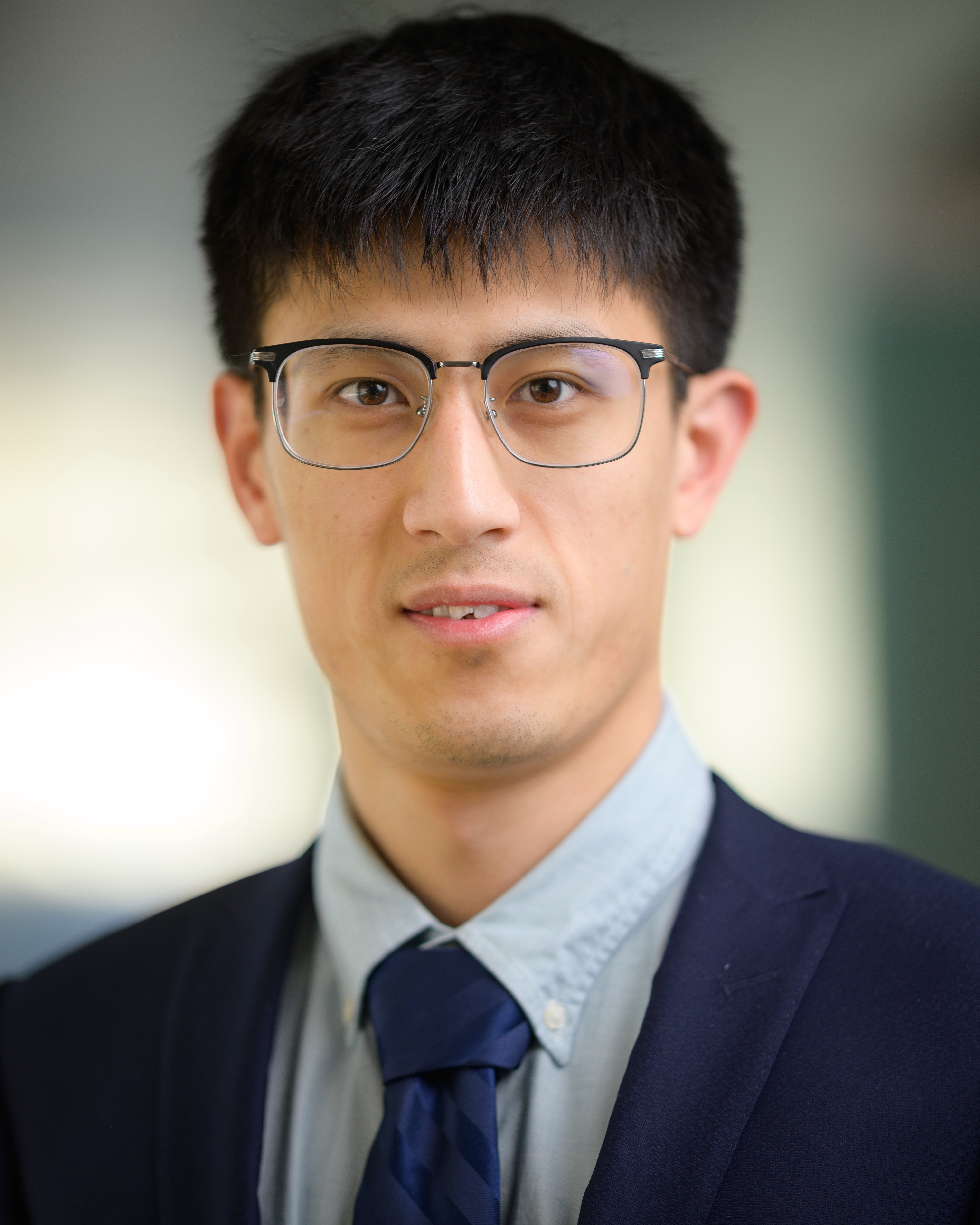}}]%
{Kaiwen Liu} received the B.S. degree in electrical and computer engineering from Shanghai Jiao Tong Univeristy, Shanghai, China in 2017, and the B.S. and the M.S. degrees in mechanical engineering from Univeristy of Michigan, Ann Arbor, MI, USA in 2017 and 2019, respectively, where he is currently pursuing Ph.D. degree in aerospace engineering. His current research interests include decision-making with human interactions and learning methods for constrained control problems.
\end{IEEEbiography}
\begin{IEEEbiography}
[{\includegraphics[width=1in,height=1.25in,clip,keepaspectratio]{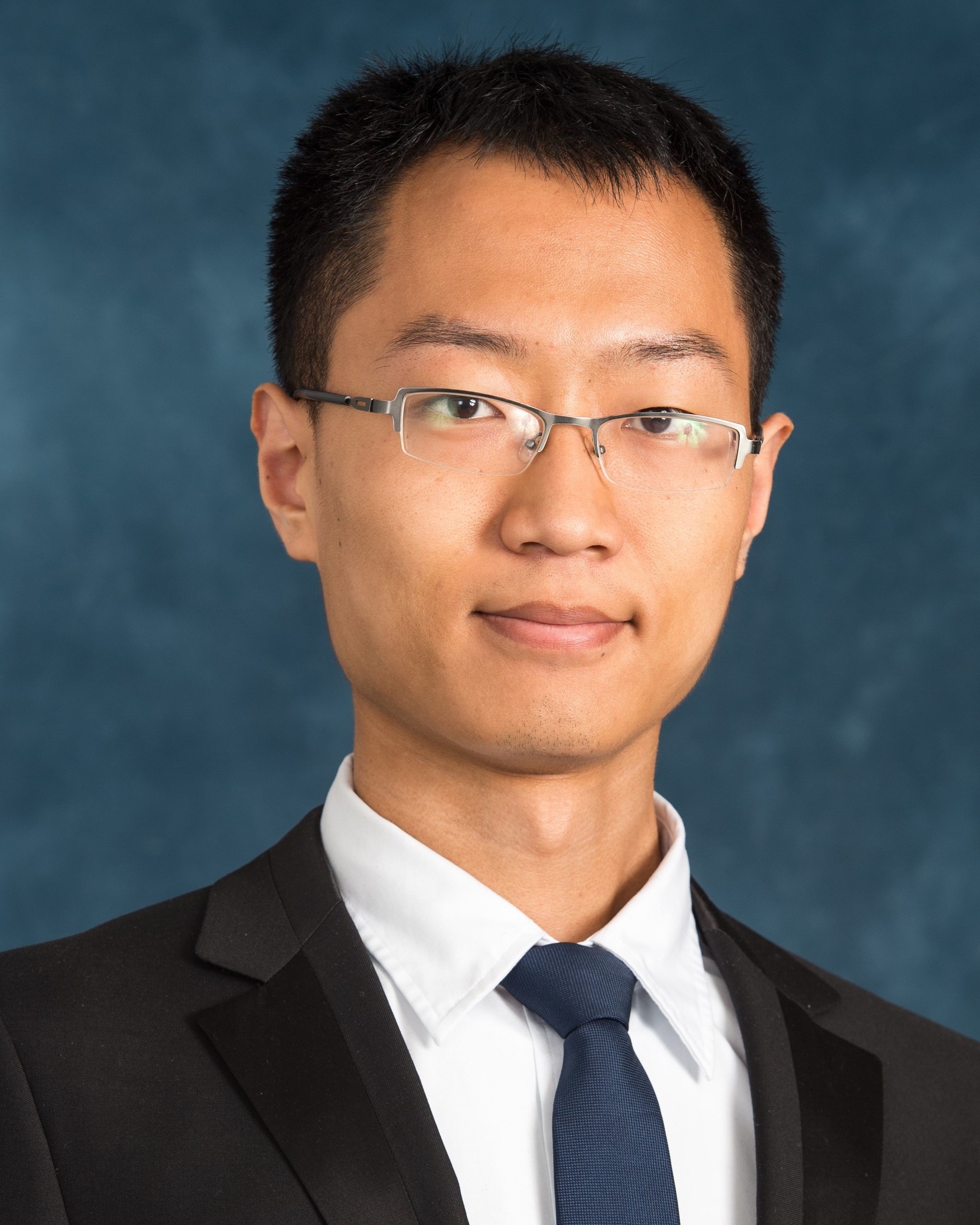}}]{Nan Li} received the Ph.D. degree in aerospace engineering and the M.S. degree in mathematics from the University of Michigan, Ann Arbor, MI, USA, in 2021 and 2020, respectively. He is currently a postdoctoral research fellow there. His research interests are in stochastic control, game theory, multi-agent and safety-critical systems, and their applications to intelligent transportation. Dr. Li was a recipient of the Professor Pierre T. Kabamba Award from the University of Michigan.
\end{IEEEbiography}

\begin{IEEEbiography}[{\includegraphics[width=1in,height=1.25in,clip,keepaspectratio]{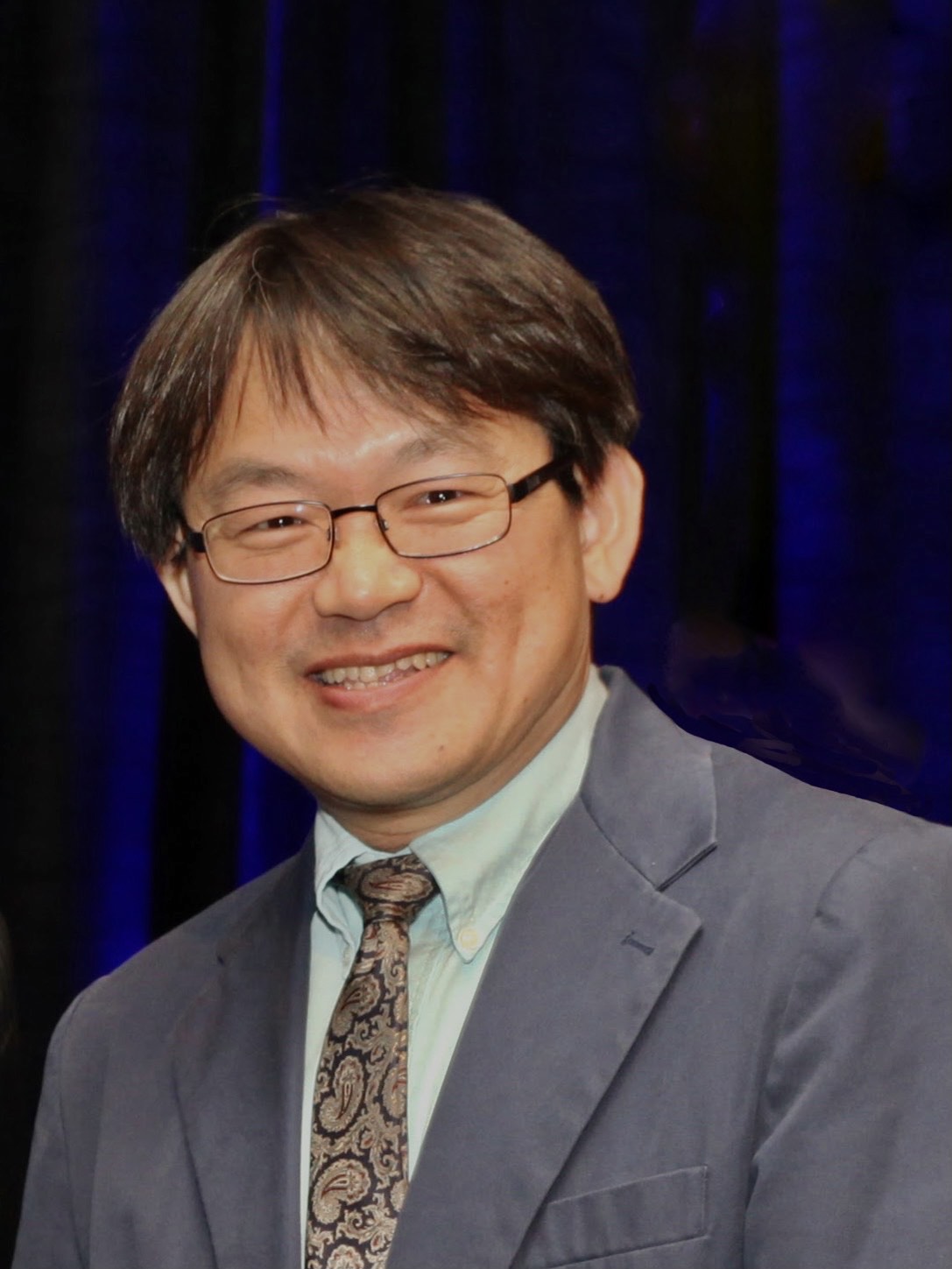}}]
{\textbf{Hongtei Eric Tseng}} received his B.S. degree from National Taiwan University, Taipei, Taiwan in 1986.  He received his M.S. and Ph.D. degrees from the University of California, Berkeley in 1991 and 1994, respectively, all in Mechanical Engineering.  In 1994, he joined Ford Motor Company. 

At Ford, he is currently a Senior Technical Leader of Controls and Automated Systems in Research and Advanced Engineering.  Many of his contributed technologies led to production vehicles implementation. His technical achievements have been recognized internally seven times with Ford’s highest technical award - the Henry Ford Technology Award, as well as externally by the American Automatic Control Council with Control Engineering Practice Award in 2013.  Eric has over 100 US patents and over 120 publications. He is an NAE Member.
\end{IEEEbiography}

\begin{IEEEbiography}
[{\includegraphics[width=1in,height=1.25in,clip,keepaspectratio]{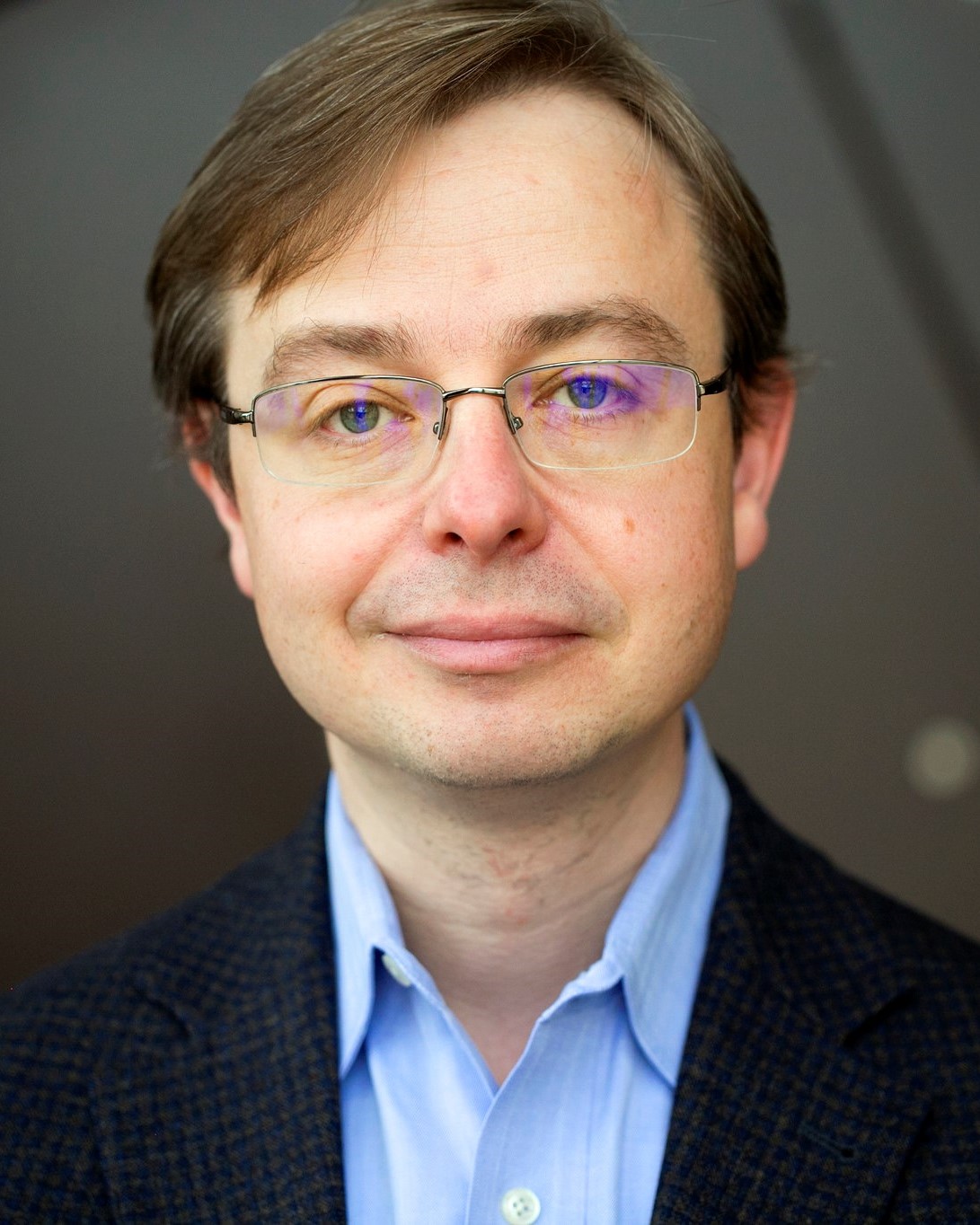}}]{Ilya Kolmanovsky} is a professor in the department of aerospace engineering at the University of Michigan, Ann Arbor, MI, USA, with research interests in control theory for systems with state and control constraints, and in control applications to aerospace and automotive systems. He received his Ph.D. degree in aerospace engineering from the University of Michigan in 1995.  Prior to joining the University of Michigan as a faculty in 2010, Kolmanovsky was with Ford Research and Advanced Engineering in Dearborn, Michigan for close to 15 years. He is a Fellow of IEEE and a Senior Editor of IEEE Transactions on Control Systems Technology.
\end{IEEEbiography}

\begin{IEEEbiography}
[{\includegraphics[width=1in,height=1.25in,clip,keepaspectratio]{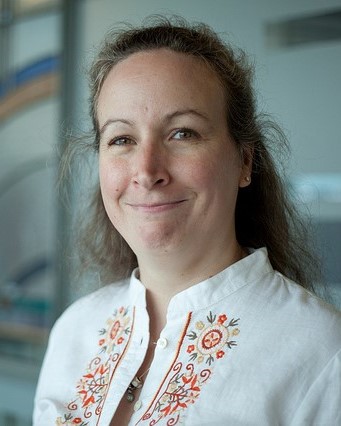}}]{Anouck Girard} received the Ph.D. degree in ocean engineering from the University of California, Berkeley, CA, USA, in 2002. She has been with the University of Michigan, Ann Arbor, MI, USA, since 2006, where she is currently a Professor of Aerospace Engineering. She has co-authored the book Fundamentals of Aerospace Navigation and Guidance (Cambridge University Press, 2014). Her current research interests include vehicle dynamics and control systems. Dr. Girard was a recipient of the Silver Shaft Teaching Award from the University of Michigan and a Best Student Paper Award from the American Society of Mechanical Engineers.
\end{IEEEbiography}
